\documentclass[prd,aps,showpacs,showkeys,eqsecnum]{revtex4}

\usepackage{amssymb}
\usepackage{amsmath}
\usepackage{latexsym}

\usepackage[dvips]{graphicx}

\newcommand{\eqb}{\begin{equation}}
\newcommand{\eqe}{\end{equation}}
\newcommand{\eqbnon}{\begin{equation*}}
\newcommand{\eqenon}{\end{equation*}}

\newcommand{\eqab}{\begin{eqnarray}}
\newcommand{\eqae}{\end{eqnarray}}
\newcommand{\eqabnon}{\begin{eqnarray*}}
\newcommand{\eqaenon}{\end{eqnarray*}}
\newcommand{\seqb}{\begin{subequations}}
\newcommand{\seqe}{\end{subequations}}

\newcommand{\Cs}{\left.C\right|_{\Sigma}}
\newcommand{\vac}{|\text{vac}\rangle}
\newcommand{\vacbra}{\langle\text{vac}|}
\newcommand{\bvac}{|\overline{\text{vac}}\rangle}
\newcommand{\bvacbra}{\langle\overline{\text{vac}}|}
\newcommand{\bu}{\bar{u}}
\newcommand{\bv}{\bar{v}}
\newcommand{\bm}{\bar{\mu}}
\newcommand{\bn}{\bar{\nu}}
\newcommand{\be}{\bar{\eta}}
\newcommand{\bc}{\bar{\chi}}
\newcommand{\tmn}[1]{\langle T_{#1} \rangle}

\newcommand{\ttmn}[1]{\langle \widetilde{T}_{#1} \rangle}
\newcommand{\tg}{\widetilde{g}}
\newcommand{\tW}{\widetilde{W}}

\newcommand{\defeq}{:=}
\newcommand{\defeqr}{=:}

\begin{document}

\title{Black hole evaporation in an expanding universe}

\author{Hiromi Saida}
\email{saida@daido-it.ac.jp}
\affiliation{Department of Physics, Daido Institute of Technology, Minami-ku, Nagoya 457-8530, Japan}

\author{Tomohiro Harada}
\email{harada@rikkyo.ac.jp}
\affiliation{Department of Physics, Rikkyo university, Toshima-ku, Tokyo 171-8501, Japan}

\author{Hideki Maeda}
\email{hideki@cecs.cl}
\affiliation{Centro de Estudios Cient\'{\i}ficos (CECS), Arturo Prat 514, Valdivia, Chile \\
Department of Physics, International Christian University, Mitaka-shi, Tokyo 181-8585, Japan}


\begin{abstract}
We calculate the quantum radiation power of black holes which are asymptotic to the Einstein-de~Sitter universe at spatial and null infinities. 
We consider two limiting mass accretion scenarios, no accretion and significant accretion. We find that the radiation power strongly depends on not only the asymptotic condition but also the mass accretion scenario. 
For the no accretion case, we consider the Einstein-Straus solution, where a black hole of constant mass resides in the dust Friedmann universe.
We find negative cosmological correction besides the expected redshift factor. 
This is given in terms of the cubic root of ratio in size of the black hole to the cosmological horizon, so that it is currently of order $10^{-5} (M/10^{6}M_{\odot})^{1/3} (t/14\, \mbox{Gyr})^{-1/3}$ but could have been significant at the formation epoch of primordial black holes. 
Due to the cosmological effects, this black hole has not settled down to an equilibrium state. 
This cosmological correction may be interpreted in an analogy with the radiation from a moving mirror in a flat spacetime. 
For the significant accretion case, we consider the Sultana-Dyer solution, where a black hole tends to increase its mass in proportion to the cosmological scale factor.
In this model, we find that the radiation power is apparently the same as the Hawking radiation from the Schwarzschild black hole of which mass is that of the growing mass at each moment. 
Hence, the energy loss rate decreases and tends to vanish as time proceeds. 
Consequently, the energy loss due to evaporation is insignificant compared to huge mass accretion onto the black hole.
Based on this model, we propose a definition of quasi-equilibrium temperature for general conformal stationary black holes.
\end{abstract}



\maketitle

\section{Introduction}
\label{sec-intro}

Exactly speaking, in our Universe there is no black hole which is asymptotically flat. 
We call black holes asymptotic to an expanding universe {\it cosmological black holes}. 
A particularly important example of such black holes is {\it primordial black holes}~\cite{ref-pbh.1} which may have formed in the early universe. 
They would play a unique role as probes into currently unknown physics in various aspects. 
The key observable phenomenon is the Hawking radiation. 
See~\cite{ref-pbh.2} and references therein. 
The effect of cosmological expansion on the dynamics of black holes has been studied~\cite{ref-pbh.3}. 
However, the cosmological effect on the Hawking radiation has not been seriously investigated yet. 
It is generally believed that the cosmological expansion will not affect the evaporation process if the cosmological horizon is much larger than the black hole horizon. 
Although this is very plausible from a physical point of view, this should be verified from a definite argument and it is also important to estimate the possible cosmological corrections. 
Moreover, this assumption might not be the case in some cosmological situations. 
For example, the black hole horizon may have been of the same order as the cosmological horizon immediately after primordial black holes formed and/or if matters around black hole continued accreting so rapidly that self-similar growth of black hole horizon might be possible~\cite{ref-pbh.3}.

The evaporation of cosmological black holes is also important from a point of view of thermodynamics. 
The {\it black hole thermodynamics}~\cite{ref-bht} offers a unique possibility of understanding the theory of gravity through the laws of thermodynamics. 
It is argued that a black hole is described as an object in thermal equilibrium (black body) with temperature $T_{\rm H}=\kappa/2 \pi$~\cite{ref-hr,ref-qft}, where $\kappa$ is the surface gravity of the event horizon, and evaporates by radiating its mass energy according to the Stefan-Boltzmann law~\cite{ref-evapo}. 
This argument has been so far established only for black holes which are asymptotic to flat, dS and AdS spacetimes at spatial and null infinities. 
See~\cite{ref-bht} for asymptotically flat case, and \cite{ref-(A)dS} for asymptotically dS and AdS cases. 
It might be reasonable that black hole thermodynamics requires the asymptotically static nature of spacetimes, because an equilibrium state corresponds to a static condition. 
However, if black holes can be regarded as thermodynamic systems even for nonstationary cases, it is natural to ask what kind of nonequilibrium states correspond to dynamical black holes. 
An interesting example is cosmological black holes. 
The spectrum of its Hawking radiation has already been studied for the case of no mass accretion~\cite{ref-spectrum}. 
However, the radiation power or luminosity has not been explicitly calculated yet.

In the present article, we calculate the power of the Hawking radiation through the quantum expectation value of stress-energy tensor in an expanding universe based on the two limiting scenarios about mass accretion.
To do this, we use two interesting models for cosmological black holes. 
One is the Einstein-Straus black hole and the other is the Sultana-Dyer black hole. 
The former has no mass accretion, while the latter has significant mass accretion. 
Our first model was raised by Einstein and Straus~\cite{ref-cosmo.bh.ES}. 
This is the exact solution of the Einstein equation with timelike dust, which is obtained by pasting a Schwarzschild spacetime with a dust Friedmann universe on a timelike hypersurface. 
This kind of models are often termed as of ``Swiss-cheese'' type. 
The existence of a black hole event horizon in the Einstein-Straus solution is guaranteed by construction and the radius of the event horizon is constant. 
All energy conditions are of course satisfied in this spacetime. 
Our second model was given by Sultana and Dyer~\cite{ref-cosmo.bh.SD,ref-cosmo.bh.MD}. 
This is obtained by a conformal transformation operated on the Schwarzschild spacetime, whose conformal factor is carefully chosen so that the metric is the exact solution with the combination of timelike and null dusts and the spacetime is asymptotic to the Einstein-de~Sitter (or flat dust Friedmann) universe at spatial and null infinities. 
Since this transformation does not affect the causal structure, the existence of an event horizon is guaranteed, although this spacetime has some trouble with energy conditions. 
This spacetime has a feature that the physical radius of the event horizon increases due to accretion, and it approaches infinity as time proceeds although its growth rate tends to be much slower than the growth rate of the cosmological horizon.

This paper is organised as follows.
Sections \ref{sec-ES} and \ref{sec-SD} are devoted for the calculation of Hawking radiation from Einstein-Straus and Sultana-Dyer black holes, respectively. Summary and discussions are given in section \ref{sec-sd}. 
Throughout this paper, we use the Planck units, $c = \hbar = G = k_B =1$.

\section{Hawking radiation from the Einstein-Straus black hole}
\label{sec-ES}

\subsection{The Einstein-Straus black hole and its formation}
\label{sec-ES.bg}

The Einstein-Straus black hole is constructed by pasting the Schwarzschild and the Friedmann solutions at a spherically symmetric timelike hypersurface $\Sigma$ (see~\cite{ref-cosmo.bh.ES} or Appendix A in~\cite{ref-spectrum}).

The Schwarzschild metric is given by
\eqb
 ds_{\rm BH}^2 = - C(R) \, dT^2 + \frac{dR^2}{C(R)} + R^2\, d\Omega^2 \, ,
\label{eq-ES.bg-bh}
\eqe
where
\eqb
 C(R) \defeq 1 - \frac{2 M}{R} \, ,
\eqe
$M$ is the mass of this black hole, $d\Omega$ is the line element of a unit two dimensional sphere, and $T$ and $R$ are the time coordinate and areal radius in the Schwarzschild coordinates, respectively. 
We can get the double null form of the metric as 
\eqb
 ds_{\rm BH}^2 = - C(R)\, dU\, dV +R^2\, d\Omega^{2} \, , 
\label{eq-ES.bg-bh.null}
\eqe
where 
\seqb
\label{eq-ES.bg-bh.tortoise.EF}
\eqab
  dR^{\ast} &\defeq& \frac{dR}{C(R)} \, , \\
  U &\defeq& T - R^{\ast} \, , \quad V \defeq T + R^{\ast} \, .
\eqae
\seqe

The Friedmann metric is given by
\eqab
 ds_{\rm F}^2 = - dt^2 + a(t)^2 \left(\, \frac{dr^2}{1 - k r^2} + r^2\, d\Omega^2 \,\right) \, ,
\label{eq-ES.bg-frw}
\eqae
where $a$ is the scale factor, $k$ is the spatial curvature, $r$ is the comoving radius and $t$ is the proper time of the comoving observer or the cosmological time. 
Hereafter we assume the open or flat Friedmann metric ($k = 0$ or $-1$) in order to guarantee the existence of future null infinity. 
The double null form of the metric is given by 
\eqb
   ds_{\rm F}^2 = - a^2 du\, dv + R^2\, d\Omega^2 \, ,
\label{eq-ES.bg-frw.null}
\eqe
where 
\seqb
\eqab
 d\eta &\defeq& \frac{dt}{a} \, , \quad d\chi \defeq \frac{dr}{\sqrt{1 - k r^2}} \, , 
\label{eq-ES.bg-frw.eta.chi} \\
 u &\defeq& \eta - \chi \, , \quad   v \defeq \eta + \chi \, .
\label{eq-ES.bg-frw.u.v}
\eqae
\seqe

The junction surface $\Sigma$ is given by a constant comoving radius
$r=r_{\Sigma}$ and located at 
\eqb
 R = a(t)\,r_{\Sigma} \, . 
\label{eq-ES.bg-js}
\eqe
The Israel junction condition~\cite{ref-jc} with no singular hypersurface reduces the continuity of the first and second fundamental forms on the junction surface $\Sigma$. 
Then, we obtain two independent equations,
\seqb
\eqab
 && \frac{d T(t)}{dt} = \frac{\sqrt{1 - k r_{\Sigma}^2}}{\Cs} \label{eq-ES.bg-jc.T} \, , \\
 && \left(\frac{\dot{a}}{a}\right)^2 + \frac{k}{a^2}
      = \frac{2 M}{r_{\Sigma}^3} \frac{1}{a^3} \label{eq-ES.bg-jc.a} \, ,
\eqae
\seqe
where $\dot{a} \defeq d a(t)/dt$ and 
\eqb
 \Cs = 1 - \frac{2 M}{a\, r_{\Sigma}} \, .
\eqe
Equation \eqref{eq-ES.bg-jc.T} gives the relation between the time coordinates $T$ and $t$. 
Equation \eqref{eq-ES.bg-jc.a} determines the time evolution of the scale factor $a(t)$ and identical with a dust Friedmann equation. 
Thus, it is required that the Friedmann metric is that of the dust Friedmann universe. 
By comparing eq.~\eqref{eq-ES.bg-jc.a} with the dust Friedmann equation,
\eqab
 \left(\frac{\dot{a}}{a}\right)^2 + \frac{k}{a^2} \label{eq^prep.bg.jc.a}
      = \frac{8 \pi}{3} \frac{\rho_{\ast}}{a^3} \, ,
\label{eq-ES.bg-dust}
\eqae
where $\rho_{\ast}$ is a constant, we find a relation between $\rho_{\ast}$ and $M$,
\eqab
 M = \frac{4}{3} \pi \, r_{\Sigma}^3 \, \rho_{\ast} \, .
\label{eq-ES.bg-mass}
\eqae
The Einstein-Straus black hole describes a cosmological black hole spacetime with no mass accretion.

\begin{figure}[t]
 \begin{center}
 \includegraphics[height=40mm]{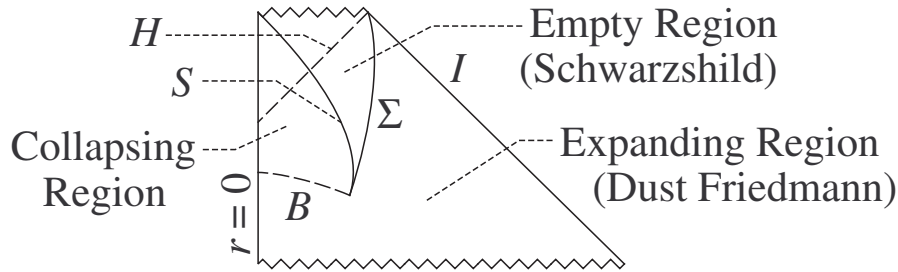}
 \end{center}
\caption{The Einstein-Straus black hole formed in an expanding universe. 
The zig-zag lines denote spacetime singularities. $H$, $I$, $B$, $S$ and $\Sigma$ denote the black hole event horizon, the null infinity, the starting spacelike surface of gravitational collapse, the surface of collapsing dust ball and the junction surface between the Schwarzschild and Friedmann spacetimes, respectively. 
The empty region surrounded by $S$, $\Sigma$ and the black hole singularity is described by the Schwarzschild solution. The collapsing region surrounded by $B$, $S$ and the regular centre is described by some regular dynamical metric. 
The expanding region surrounded by $B$, $\Sigma$, $I$, the regular centre and the big-bang singularity is described by the Friedmann solution with dust.}
\label{fig-1}
\end{figure}

To calculate the power of Hawking radiation from the Einstein-Straus black hole, we need to specify how it has formed in the expanding universe. 
Here we assume that an overdense region of which comoving radius is $r_{\Sigma}$ and mass $M$ begins to contract at the moment $t=t_{\rm B}$ and collapse to form a black hole.
The Penrose diagram of this gravitational collapse is shown in fig.~\ref{fig-1}. 
In this figure, $H$, $I$, $B$, $S$ denote the event horizon, the null infinity, the spacelike hypersurface $t=t_{\rm B}$ for $r<r_{\Sigma}$ and the surface of the collapsing dust ball, respectively. 
We can assume that the collapsing region surrounded by $B$, $S$ and the regular centre is described by some regular dynamical metric,
\eqab
 ds_{\rm col}^2 = A(\tau,\lambda) \left( - d\tau^2 + d\lambda^2 \right)
            + R(\tau,\lambda)^2 d\Omega^2 \, ,
\label{eq-ES.bg-collapse}
\eqae
where $\tau$ and $\lambda$ are respectively appropriate temporal and radial coordinates. 
We can set that $\lambda=0$ corresponds to the regular centre. 
The double null form of the metric is given by 
\eqb
   ds_{\rm col}^2 = - A\, d\alpha\, d\beta +R^{2}d\Omega^{2} \, ,
\label{eq-ES.bg-collapse.null}
\eqe
in the collapsing region, where 
\eqb
\alpha \defeq \tau-\lambda \, , \quad \beta \defeq \tau + \lambda \, .
\label{eq-ES.bg-collapse.alpha.beta}
\eqe

As assumed above, the starting surface $B$ of collapse is given by a spacelike hypersurface in the expanding region,
\eqb
 t = t_{\rm B}(r) \quad \mbox{and} \quad 0\le r \le r_{\Sigma} \, .
\eqe
This surface is also described in the collapsing region by 
\eqb
\tau = \tau_{\rm B}(r) \quad \mbox{and} \quad \lambda = \lambda_{\rm B}(r) \, ,
\label{eq-ES.bg-B}
\eqe
using the coordinates in the collapsing region.

\subsection{Redshift and Hawking radiation}
\label{sec-ES.redshift}

We introduce a matter field $\phi$ which describes quantum radiation from a black hole. 
For simplicity, let $\phi$ be a massless scalar field with minimal coupling, which satisfies the Klein-Gordon equation, $\Box\phi = 0$. 
In manipulating quantum field theory in curved spacetimes, especially on black hole spacetimes, we need to estimate the redshift.

\begin{figure}[t]
 \begin{center}
 \includegraphics[height=40mm]{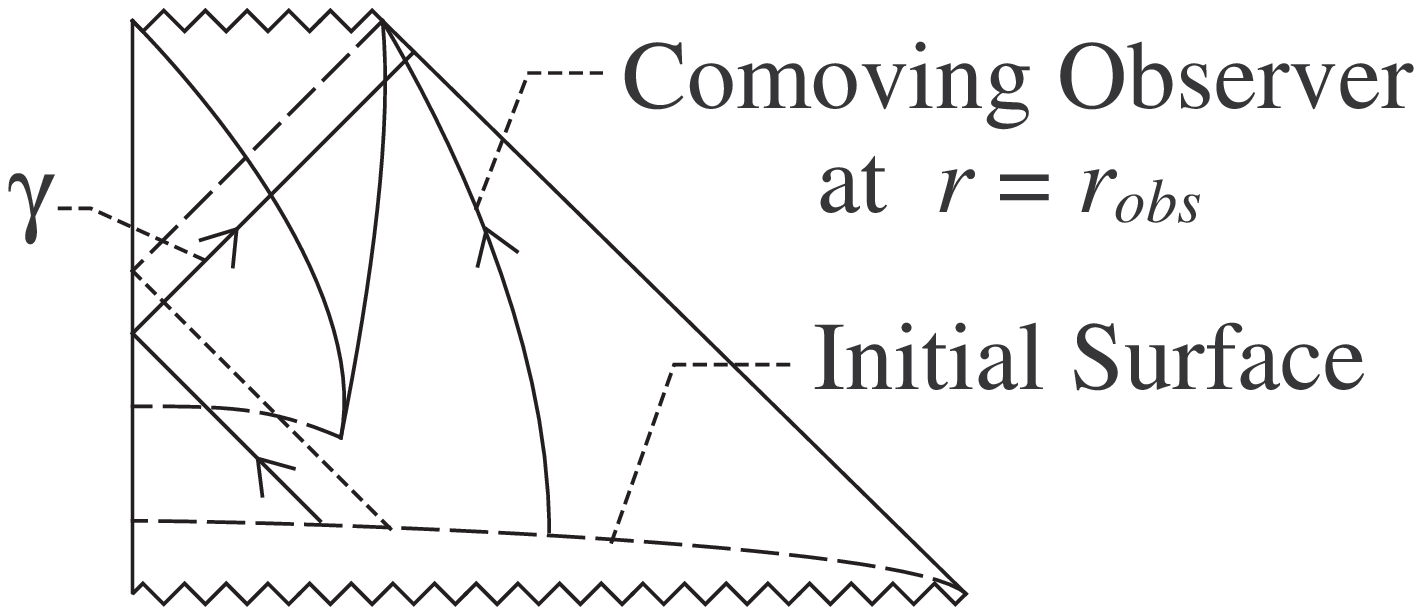} \\
 \includegraphics[height=40mm]{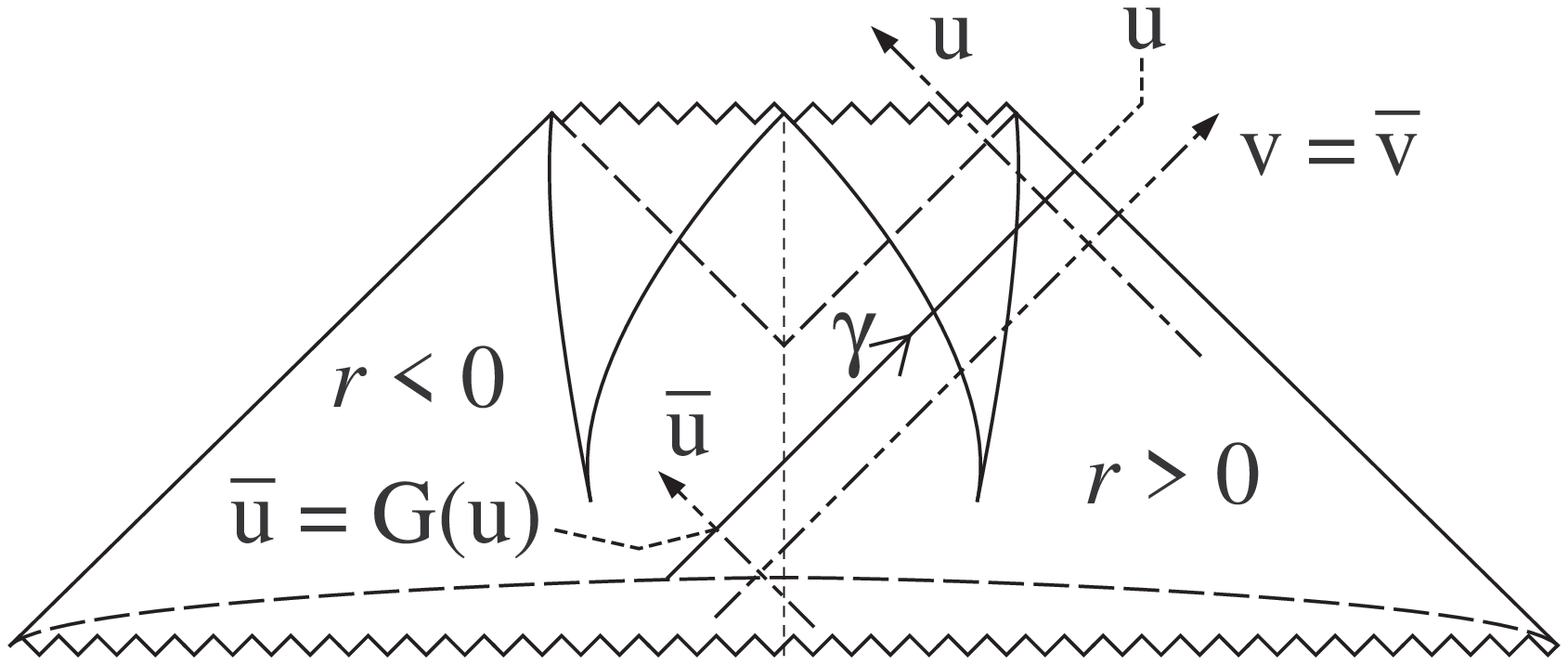}
 \end{center}
\caption{The Penrose diagrams of the cosmological black hole with no matter accretion. 
{\bf Upper panel}: The initial surface and the comoving observer at $r_{\rm obs} > r_{\Sigma}$ are shown.  
The null geodesic $\gamma$ passes near the event horizon. 
A late-time observer detects a scalar wave which has propagated along $\gamma$. 
{\bf Lower panel}: The Penrose diagram of gravitational collapse extended to the negative radius region. 
This extension makes it easier to calculate $\tmn{\mu \nu}$.}
\label{fig-2}
\end{figure}

The wave mode of $\phi$ propagates along a null geodesic $\gamma$ passing near the event horizon. 
As shown in the upper panel in fig.~\ref{fig-2}, this mode is ingoing at the initial surface and becomes outgoing after passing through the centre. 
The function $\bv = G(u)$ relates the ingoing null coordinate $\bv$ of $\gamma$ at the initial surface and the outgoing null coordinate $u$ of $\gamma$ at late times. 
Here note that, although both the initial surface and the spacetime region where a distant observer is are given by the Friedmann solution, we need to distinguish the null coordinates by $(\bu,\bv)$ at the initial surface and by $(u,v)$ for the observer because their values are different from each other in this construction. 
The function $G(u)$ is obtained by the junction of null coordinates at the intersections of $\gamma$ with the surfaces $B$, $S$ and $\Sigma$.

The junction at $B$ gives a relation between $\bv$ and $\beta$ as $\bv = \bv(\beta)$. 
Since we can assume that the gravitational collapse begins with sufficient smoothness, $\bv = \bv(\beta)$ is $C^{1}$ for the relevant ingoing null rays, i.e.,
\eqb
 \bv = \bv_1 \, \beta + \bv_{0} \, ,
\label{eq-ES.redshift-B}
\eqe
where $\bv_{0}$ and $\bv_{1}(>0)$ are constants
\footnote{A similar discussion is already in \S 8.1 in~\cite{ref-qft}.}. 
The reflection of $\gamma$ at the regular centre is given by a simple replacement of $\beta$ by $\alpha$ as $\bv = \bv(\alpha)$. 
This expresses the redshift along $\gamma$ from the initial surface to the collapsing region.

The junction at $S$ gives a relation between $\alpha$ and $U$ as $\alpha = \alpha(U)$. 
This junction is completely the same as one gets for an asymptotically flat black hole (see~\cite{ref-hr} and \S8.1 in~\cite{ref-qft}). 
Hence, we do not show the details of the calculation but only quote the result:
\eqb
 \alpha = \alpha_{1} \exp\left( - \kappa\, U \right) + \alpha_{0} \, ,
\label{eq-ES.redshift-S}
\eqe
where $\alpha_{0}$ and $\alpha_{1}(>0)$ are constants and 
\eqb
\kappa \defeq \left.\frac{1}{2}\frac{dC}{dR}\right|_{R=2M} = \frac{1}{4 M}.
\eqe 
This expresses the redshift from the collapsing region to the empty region.

The junction on $\Sigma$ gives a relation between $U$ and $u$ as $U = U(u)$. 
To obtain this, we need to consider the junction of metrics, $\left.ds_{\rm BH}^2\right|_{\Sigma} = \left.ds_{\rm F}^2\right|_{\Sigma}$. 
Since $\Sigma$ is the timelike hypersurface given by $r=r_{\rm \Sigma}$ in the comoving coordinates, the null coordinates on $\Sigma$ are regarded as functions of $t$. 
Furthermore, from the relation \eqref{eq-ES.bg-frw.eta.chi}, these null coordinates are instead regarded as functions of $\eta$. 
Then the junction of the null coordinates on $\Sigma$ can be discussed using the partial derivative with respect to $\eta$, i.e., the junction of metrics gives the following relation on $\Sigma$,
\eqb
 C\, V_{,\eta}\, U_{,\eta} = a^2 \, .
\label{eq-ES.redshift-jc.Sigma}
\eqe
On the other hand, from eqs.~\eqref{eq-ES.bg-bh.tortoise.EF} and \eqref{eq-ES.bg-js}, we get on $\Sigma$
\eqb
 V_{,\eta}-U_{,\eta} = 2 \frac{r_{\Sigma}\,a'}{C} \, ,
\eqe
where the prime $'$ denotes the argument differential, $a' \defeq da(\eta)/d\eta$. 
Therefore, eq.~\eqref{eq-ES.redshift-jc.Sigma} becomes a quadratic equation
for $U_{,\eta}$ on $\Sigma$. 
We get the positive root and obtain $dU/du$ as
\eqb
 \frac{dU}{du} 
  = \left.\frac{U_{,\eta}}{u_{,\eta}} \right|_{\Sigma}
  = \left.\frac{1}{C}
      \left( - r_{\Sigma}\,a'
             + \sqrt{{r_{\Sigma}^2} a'^{2} + a^2 C} \right) \right|_{\Sigma} \, .
\label{eq-ES.redshift-Sigma}
\eqe
This expresses the redshift along $\gamma$ from the empty region to a comoving observer in the expanding region.

Thus, we obtain the function $G(u)$ by combining three functions and the reflection at the centre.
\eqb
 \bv = G(u) = \bv_{1}' \exp\left[ - \kappa\, U(u) \right] + \bv_{0}' \, ,
\eqe
where $\bv'_{0}$ and $\bv'_{1}(>0)$ are constants.
Furthermore, for later convenience, we will extend the background to include the negative radial coordinate region. 
The extended Penrose diagram is shown in the lower panel in fig.~\ref{fig-2}. 
Then we do not need to consider the reflection of $\gamma$ at the centre. 
By this virtual extension, eq.~\eqref{eq-ES.redshift-B} becomes $\bu = \bu_{1}  \alpha + \bu'_{0}$, where $\bu'_{0}$ and $\bu_{1}(>0)$ are constants. 
Hence, the redshift in the extended background spacetime is given by the same function $G(u)$ obtained above with replacing $\bv$ by $\bu$ in the left-hand side,
\eqb
 \bu = G(u)  = \bu_1 \exp\left[ - \kappa\, U(u) \right] + \bu_0 \, ,
\label{eq-ES.redshift-redshift}
\eqe
where $\bu_0$ is a constant. 
Note that this function \eqref{eq-ES.redshift-redshift} also applies to the dimensionally reduced spacetime introduced below.

\subsection{Hawking radiation from cosmological black hole with no accretion}
\label{sec-ES.stress}

The quantum expectation value must be renormalised. 
The regularisation technique in two dimensions has been well established. 
Appendix summarises the calculation of the vacuum expectation value of stress-energy tensor $\vacbra T_{\mu \mu} \vac$ and its renormalised value $\tmn{\mu \nu}$, where $\vac$ is an appropriate initial vacuum state. 
For simplicity, we here reduce the gravitational collapse spacetime described by fig.~\ref{fig-2} to a two dimensional one by cutting out the two dimensional angular part from metrics \eqref{eq-ES.bg-bh}, \eqref{eq-ES.bg-frw} and \eqref{eq-ES.bg-collapse}. 
It has been known that two dimensional Schwarzschild black hole gives a qualitatively correct power of Hawking radiation in four dimensions, since the so-called {\it grey body factor} in the Hawking radiation disappears due to the absence of curvature scattering of matter fields in two dimensions.
The curvature scattering of matter fields also does not occur in two dimensional Einstein-Straus spacetime. 
We can expect that two dimensional Einstein-Straus black hole gives a qualitatively correct radiation power.

The thermal radiation in asymptotically flat black hole spacetimes has been obtained under the following three procedures~\cite{ref-hr,ref-qft}; 
to neglect the curvature scattering, to define an initial vacuum state on a spacelike hypersurface before the black hole formation, and to observe particles at sufficiently late times. 
The first is automatically done if we work in two dimensions. 
The second implies that, in the Heisenberg picture, the quantum state which has been vacuum initially is no longer vacuum after the gravitational collapse. 
The third implies that a wave mode detected by a distant observer should pass the neighborhood of the event horizon and hence it has been strongly redshifted before it is observed. 
This means that it has been of very high frequency near the event horizon, and the geometrical optics approximation is valid, being consistent with neglecting the curvature scattering even for four dimensional case.

To calculate the Hawking radiation from the Einstein-Straus black hole, we adopt the vacuum state $\bvac$ associated with a comoving observer at the initial surface as a physical initial vacuum state and calculate $\tmn{\mu \nu}$ for a distant comoving observer at sufficiently late times. 
To be precise, $\bvac$ is defined by the quantisation of $\phi$ using the normal modes obtained in the coordinate system of eq.~\eqref{eq-ES.bg-frw.u.v} on the initial surface, and the components of $\tmn{\mu \nu}$ is calculated in the same coordinates. 
Moreover, in two dimensions there is no genuine cosmological particle creation for a massless scalar field (see \S3.4 in~\cite{ref-qft} for example), and hence $\tmn{\mu \nu}$ expresses purely the Hawking radiation from cosmological black holes.

First, we calculate $\tmn{\bm \bn}^{\rm (in)}$ at the initial surface using eqs.~\eqref{eq-app.two-stress.1} and \eqref{eq-app.two-stress.2} given in Appendix. 
The metric suitable for this purpose is given by $ds_{\rm F}^2 = - a^2 d\bu d\bv$, for which
\eqb
{\cal R} = - \frac{ 2[(a')^2 - a a'' ]}{a^4}
\eqe
in eq.~\eqref{eq-app.two-stress.1} and $D = a^2$ in eq.~\eqref{eq-app.two-stress.2}. 
Therefore we obtain
\seqb
\eqab
 \tmn{\bu \bu}^{\rm (in)}
 &=& \tmn{\bv \bv}^{\rm (in)} = -\frac{2 (a')^2 - a\,a''}{48 \pi\,a^2} \, , \\
 \tmn{\bu \bv}^{\rm (in)}
 &=& \frac{(a')^2 - a\,a''}{48 \pi\,a^2} \, .
\eqae
\seqe
The observed power for a comoving observer on the initial surface is proportional to $\be$-$\bc$ component of $\tmn{\bm \bn}^{\rm (in)}$. 
By coordinate transformation from $(\bu,\bv)$ to $(\be,\bc)$, we find
\eqab
 \tmn{\be \bc}^{\rm (in)} = 0 \, .
\label{eq-ES.stress-power.initial}
\eqae
This indicates that no energy flux is observed at the initial surface.

Next we calculate $\tmn{\mu \nu}^{\rm (obs)}$ measured by a comoving observer at late times. 
The metric suitable for this purpose is $ds_{\rm F}^2 = - a^2 du\, dv$. 
This gives the same ${\cal R}$ as for $\tmn{\bm \bn}^{\rm (in)}$. 
In order to find the function $D$ in eq.~\eqref{eq-app.two-stress.2}, null coordinates $(u,v)$ of a late-time observer should be expressed in terms of null coordinates $(\bu,\bv)$ with which the initial vacuum state $\bvac$ is defined. 
The relation between $(u,v)$ and $(\bu,\bv)$ reflects the time evolution of the background spacetime in between and is given by the redshift along outgoing and ingoing null geodesics which connect the late-time observer and the initial surface. 
For the extended background spacetime (lower panel in fig.~\ref{fig-2}), the relation between $u$ and $\bu$ is given by the redshift \eqref{eq-ES.redshift-redshift} of outgoing null geodesic $\gamma$. 
The relation between $v$ and $\bv$ is a simple one $v = \bv$, because the relevant ingoing null geodesic lies in the expanding region without passing any surfaces $B$, $S$ and $\Sigma$ during propagating from the initial surface to the late-time observer. 
Consequently, $D$ which gives the metric at late times as 
$ds_{\rm F}^2 = - D d\bu\, d\bv$ is obtained as
\eqb
 D = a^{2}\frac{du}{d\bar{u}} = \frac{a^2}{G'} \, ,
\eqe
where the coordinate transformation \eqref{eq-ES.bg-frw.u.v} is used and $G' \defeq dG(u)/du$. 
This gives
\seqb
\eqab
  D_{,\bu}    &=& \frac{a\,a'}{G'^2} - \frac{a^2 G''}{G'^3} \, , \\
  D_{,\bu\bu} &=& \frac{(a\, a')'}{2 G'^3}
                 - 3 \frac{a\,a' G''}{G'^4} - \frac{a^2 G'''}{G'^4}
                 + 3 \frac{a^2 G''^2}{G'^5} \, , \\
  D_{,\bv}    &=& \frac{a\,a'}{G'} \, , \\
  D_{,\bv\bv} &=& \frac{(a\,a')'}{2 G'} \, ,
\eqae
\seqe
where $a_{,\bu} = (d u/d\bu) (\partial\eta/\partial u) a' = a'/2 G'$ and $a_{,\bv} = (d v/d\bv) (\partial\eta/\partial v) a' = a'/2$ are used. 
Hence eq.~\eqref{eq-app.two-stress.1} gives
\seqb
\eqab
 \tmn{\bu \bu}^{\rm (obs)}
 &=& \frac{1}{24 \pi}
     \left[ \frac{3}{2} \left(\frac{G''}{G'}\right)^2
          - \frac{G'''}{G'} \right] + \tmn{\bv \bv} \, , \\
 \tmn{\bv \bv}^{\rm (obs)}
 &=& - \frac{2 (a')^2 - a\,a''}{48 \pi\,a^2} \, ,\\
 \tmn{\bu \bv}^{\rm (obs)}
 &=& \frac{(a')^2 - a\,a''}{48 \pi\,a^2} \, .
\eqae
\seqe
The observed power $P_{\rm obs}$ is given by the tetrad component $\langle T_{(\eta)}^{(\chi)}\rangle^{\rm (obs)} $. We obtain $P_{\rm obs}$ from the above calculations,
\eqb
 P_{\rm obs} \defeq
     \langle T_{(\eta)}^{(\chi)}\rangle^{\rm (obs)}
   = -\frac{1}{a_{0}^2} \tmn{\eta \chi}^{\rm (obs)}
   = \frac{1}{24 \pi\, a_{0}^2}
     \left[ \frac{3}{2} \left(\frac{G''}{G'}\right)^2
          - \frac{G'''}{G'} \right] \, ,
\eqe
where $a_{0} \defeq a(\eta_{0})$ and $\eta_{0}$ is the conformal time at the moment of observation. 
By substituting the expression \eqref{eq-ES.redshift-redshift} for $G(u)$ into $P_{\rm obs}$,
\eqb
 P_{\rm obs}
   = \frac{\kappa^2}{48 \pi} \left(\frac{U'}{a_{0}}\right)^2
   + \frac{1}{24 \pi\, a_{0}^2}
     \left[ \frac{3}{2} \left(\frac{U''}{U'}\right)^2
	  - \frac{U'''}{U'} \right] \, ,
\label{eq-ES.stress-power}
\eqe
where $U' \defeq dU/du$ is given by eq.~\eqref{eq-ES.redshift-Sigma}.

By comparing eq.~\eqref{eq-ES.stress-power} with eq.~\eqref{eq-ES.stress-power.initial}, it is obvious that the quantum creation of energy flow occurs due to the 
forming black hole. 
Here, recall that the power of the Hawking radiation $P_{\rm H(2D)}$ in an asymptotically flat two dimensional black hole is
\eqab
 P_{\rm H(2D)}
  = \frac{1}{2 \pi} \int_0^{\infty} 
d\omega \frac{\omega}{\exp(2 \pi \omega/\kappa) - 1}
  = \frac{\kappa^2}{48 \pi} \, .
\label{eq-ES.stress-power.2D.flat}
\eqae
Comparing  $P_{\rm obs}$ with $P_{\rm H(2D)}$, we find that the factor $\left(U'/a_{0}\right)^2$ in the first term and the whole of the second term in eq.~\eqref{eq-ES.stress-power} are the effects of cosmological expansion.

Here we should recall that our calculation is performed on a two dimensional background spacetime. 
That is, in calculating $P_{\rm H(2D)}$ in eq.~\eqref{eq-ES.stress-power.2D.flat}, the state density $N/2\pi$ at energy level $\omega$ is appropriate to two dimensional case (one spatial dimension), where $N$ is the effective degrees of freedom and $N = 1$ for a scalar field. 
Therefore the numerical factor in eq.~\eqref{eq-ES.stress-power} will be valid only for two dimensional case. 
However, we expect that eq.~\eqref{eq-ES.stress-power} qualitatively correct even for four dimensional case if we neglect the curvature scattering and the cosmological particle creation.

\subsection{Application to two dimensional Einstein-Straus black hole}
\label{sec-ES.power}

Here we apply eq.~\eqref{eq-ES.stress-power} to our collapse model shown in fig.~\ref{fig-2}.
In the following, we assume $k = 0$ for simplicity. 
Then, from eq.~\eqref{eq-ES.bg-jc.a} and the relation $a\, d\eta = dt$, the scale factor becomes
\eqab
 a = \left(\frac{t}{t_{\rm in}}\right)^{2/3} = \left(\frac{\eta}{\eta_{\rm in}}\right)^2 \, ,
\label{eq-ES.power-a}
\eqae
where $\eta_{\rm in} = 3\, t_{\rm in}$, and $t_{\rm in}$ and $\eta_{\rm in}$ are respectively the cosmological and conformal times at the initial surface. 
We normalise the scale factor at the initial surface. 
Furthermore, the Friedmann equation \eqref{eq-ES.bg-dust} relates $t_{\rm in}$ and $\eta_{\rm in}$ with $\rho_{\ast}$,
\eqb
 \rho_{\ast} = \frac{3}{2 \pi \eta_{\rm in}^2} \, .
\eqe
Then eq.~\eqref{eq-ES.bg-mass} gives
\eqb
 M = \frac{2\,r_{\Sigma}^3}{\eta_{\rm in}^2} \, .
\label{eq-ES.power-mass}
\eqe
Substituting eq.~\eqref{eq-ES.power-a} into the right-hand side of eq.~\eqref{eq-ES.redshift-Sigma}, we obtain
\eqb
 U' = \left.\frac{a}{F}\right|_{\Sigma} \, ,
\label{eq-ES.power-Uu}
\eqe
where
\eqb
 F \defeq 1 + \frac{2\, r_{\Sigma}}{\eta} \, ,
\eqe
where eqs.~\eqref{eq-ES.power-mass} and $\eta_{\rm in} = 3\, t_{\rm in}$ are used. Hence, substituting eq.~\eqref{eq-ES.power-Uu} into eq.~\eqref{eq-ES.stress-power}, we obtain
\eqb
 P_{\rm obs}
 = \left(\frac{a_{\rm ret}}{a_{0}}\right)^2 
   \frac{\kappa^2}{48 \pi} \frac{1}{F(\eta_{\rm ret})^2}
 + \left(\frac{a_{\rm ret}}{a_{0}}\right)^2 \frac{1}{24\pi\, a_{\rm ret}^2}
   \left[ \frac{3}{2}\left(\frac{a^{\prime}}{a}\right)^2 - \frac{a^{\prime\prime}}{a}
          - \frac{a^{\prime}}{a}\frac{F^{\prime}}{F} + \frac{F^{\prime\prime}}{F}
          - \frac{1}{2}\left(\frac{F^{\prime}}{F}\right)^2 \right]_{\rm ret} \, ,
\label{eq-ES.power-power.2D.obs}
\eqe
where $Q_{\rm ret}$ denotes the evaluation of $Q$ at $\eta = \eta_{\rm ret} \defeq \eta_{0}-(r_{\rm obs}-r_{\Sigma})$ when the ray $\gamma$ intersects $\Sigma$.
This $P_{\rm obs}$ is regarded as a function of the cosmological time $t_{0}$ of the observer by using eq.~\eqref{eq-ES.power-a}, which gives
\eqab
 t_{\rm ret} = \frac{t_{\rm in}}{\eta_{\rm in}^3}\, \eta_{\rm ret}^3
     = \frac{t_{\rm in}}{\eta_{\rm in}^3}\,
       \left[ \frac{\eta_{\rm in}}{t_{\rm in}^{1/3}}\, t_{0}^{1/3}
            - \left( r_{\rm obs} - r_{\Sigma} \right) \right]^3
     = \left[ t_{0}^{1/3} - \left(\frac{M}{6}\right)^{1/3}            
              \left(\frac{r_{\rm obs}}{r_{\Sigma}}-1\right) \right]^3 \, ,
\label{eq-ES.power-time}
\eqae
where eq.~\eqref{eq-ES.power-mass} is used in the last equality. 

Furthermore the observed power $P_{\rm obs}$ can be expressed in a more convenient form. 
Using eq.~\eqref{eq-ES.power-mass} and the Hubble parameter
\eqb
 H \defeq \frac{a'}{a^{2}} = \frac{2}{\eta\,a} \, ,
\label{eq-ES.power-H}
\eqe
we can get 
\eqb
 \frac{2r_{\Sigma}}{\eta_{\rm ret}} = (2 M H_{\rm ret})^{1/3} = \epsilon^{1/3},
\label{eq-ES.power-epsilon}
\eqe
where $\epsilon \defeq 2 M H_{\rm ret}$ is the ratio of the black hole horizon radius to the Hubble horizon radius when $\gamma$ intersects $\Sigma$. 
Furthermore, we observe the cosmological redshift $z$ of the photon emitted from $\Sigma$,
\eqb
 1 + z \defeq \frac{a_{0}}{a_{\rm ret}} = \left(\frac{\eta_{0}}{\eta_{\rm ret}}\right)^2 \, .
\label{eq-ES.power-z}
\eqe
This $z$ will be regarded as the redshift of the host galaxy of the black hole, and the ratio $\epsilon$ can be expressed as $\epsilon = 2MH_{0}(1+z)^{3/2}$, where $H_0$ is the present Hubble parameter. 
It is very natural that the cosmological correction is given in terms of the ratio $\epsilon$. 
Using this ratio, we can express $P_{\rm obs}$ in eq.~\eqref{eq-ES.power-power.2D.obs} simply as
\eqb
 P_{\rm obs} =
  \frac{\kappa^{2}}{48 \pi\, (1+z)^{2}} \,
  \left[ (1 + \epsilon^{1/3})^{-2}
       + 8 \epsilon^{2}
         \left\{ 1 + \frac{\epsilon^{1/3}}{1+\epsilon^{1/3}}
                   - \frac{1}{8} \left( \frac{\epsilon^{1/3}}{1+\epsilon^{1/3}} \right)^{2}
         \right\}\right] \, ,
\label{eq-ES.power-power.2D.obs.2}
\eqe
where $\kappa=1/(4M)$.

Note that the observed power $P_{\rm obs}$ is not intrinsic but cosmologically redshifted.
The intrinsic power $P_{\rm ES(2D)}$ is then given by
\eqb
 P_{\rm ES(2D)} \defeq (1+z)^2 \, P_{\rm obs}
 = \frac{\kappa^{2}}{48 \pi} \,
   \left[ (1 + \epsilon^{1/3})^{-2}
        + 8 \epsilon^{2}
          \left\{ 1 + \frac{\epsilon^{1/3}}{1+\epsilon^{1/3}}
                    - \frac{1}{8} \left( \frac{\epsilon^{1/3}}{1+\epsilon^{1/3}} \right)^{2}
          \right\}\right] \, .
\label{eq-ES.power-power.2D}
\eqe
The evaporation should be described by this intrinsic power. 
Up to $O(\epsilon^{1/3})$, we get $P_{\rm ES(2D)} \simeq P_{\rm H(2D)}\,(1 - 2 \epsilon^{1/3})$. 
This implies that the intrinsic power is suppressed by the cosmological expansion. The physical interpretation of this effect is proposed in section \ref{sec-sd}.

We are interested in two distinct limits from a physical point of view. 
In the first, the event horizon is much smaller than the Hubble horizon at present.
This corresponds to the limit $\epsilon \to 0$ with keeping $z$ constant, and we obtain $P_{\rm ES(2D)} \to P_{\rm H(2D)}$ and $P_{\rm obs} \to P_{\rm H(2D)}/(1+z)^2$. In the second, we consider a very late phase of the cosmological evolution, i.e., $\eta_0 \to \infty$. 
This corresponds to the limit $\epsilon \to 0$ and $z \to 0$ simultaneously as seen from eqs.~\eqref{eq-ES.power-H} and \eqref{eq-ES.power-z}. 
Then we obtain $P_{\rm ES(2D)} \to P_{\rm H(2D)}$ and $P_{\rm obs} \to P_{\rm H(2D)}$.

In black hole thermodynamics~\cite{ref-bht}, the Schwarzschild black hole is regarded as in thermal equilibrium, and the temperature $T_{\rm H}$ is assigned to the black hole (zeroth law). 
This temperature is given by $T_{\rm H} = \kappa/2 \pi = 1/(8 \pi M)$ which satisfies the Stefan-Boltzmann law in two dimensions $P_{\rm H(2D)} = (\pi/12)\, T_{\rm H}^2$ as seen from eq.~\eqref{eq-ES.stress-power.2D.flat}. 
Then one might also want to assign the temperature $T_{\rm H}= 1/(8 \pi M)$ also to the Einstein-Straus black hole.
However, the radiation power $P_{\rm ES(2D)}$ deviates from the Stefan-Boltzmann law due to the correction term of $O(\epsilon^{1/3})$.
This suggests that the Einstein-Straus black hole deviates from thermal equilibrium in a finite cosmological time.
Only in the limit $\eta_{0}\to \infty$, this black hole settles down to thermal equilibrium.

\subsection{Evaporation of the Einstein-Straus black hole in four dimensions}
\label{sec-ES.evapo}

The power $P_{\rm H(2D)} = \kappa^2/48 \pi$ is obtained for asymptotically flat two dimensional black holes. 
Therefore, we simply replace the factor $\kappa^2/48 \pi$ in eq.~\eqref{eq-ES.power-power.2D} by the four dimensional counterpart $P_{\rm H(4D)}$. 
This $P_{\rm H(4D)}$ is given by the Stefan-Boltzmann law in four dimensions,
\eqb
 P_{\rm H(4D)} = \sigma\, T_{\rm H}^4\, A_{\rm H} = \frac{N}{30720 \pi\, M^2} \, ,
\label{eq-ES.evapo-power.4D.flat}
\eqe
where $A_{\rm H} = 4 \pi (2 M)^2$ is the spatial area of the event horizon and $\sigma = N \pi^2/120$ is the Stefan-Boltzmann constant for the massless matter field with the effective degrees of freedom $N$. 
Here $N$ is given by
\eqb 
 N \defeq n_{\rm b} + \frac{7}{8} n_{\rm f} \, ,
\label{eq-ES.evapo-N}
\eqe
where $n_{\rm b}$ and $n_{\rm f}$ are the numbers of helicities of massless bosonic and fermionic fields, respectively, and the factor $7/8$ comes from the difference of statistics of fermions from bosons (see for example~\cite{ref-evapo} for derivation). 
Then it is appropriate to estimate the order of $N$ by the standard particles (inner states of quarks, leptons and gauge particles of four fundamental interactions), $N \simeq 100$ if the black hole temperature is lower than $\simeq $ 1 TeV.
Next we consider about the correction terms in square brackets in eq.~\eqref{eq-ES.power-power.2D}. These terms come from the factors $U'$, $U''$ and $U'''$ in eq.~\eqref{eq-ES.stress-power}. 
Here recall that the function $G(u)$ in eq.~\eqref{eq-ES.redshift-redshift} is valid even for four dimensions. 
Hence we can expect that the same correction terms appear as well for four dimensional case. From the above consideration, we expect that the four dimensional intrinsic power $P_{\rm ES(4D)}$ is given by
\eqb
 P_{\rm ES(4D)} = \frac{N}{30720 \pi\, M^2}
   \left[ (1 + \epsilon^{1/3})^{-2}
        + 8 \epsilon^{2}
          \left\{ 1 + \frac{\epsilon^{1/3}}{1+\epsilon^{1/3}}
                    - \frac{1}{8} \left( \frac{\epsilon^{1/3}}{1+\epsilon^{1/3}} \right)^{2}
          \right\}\right] \, .
\label{eq-ES.evapo-power.4D}
\eqe

Finally we estimate the evaporation time of the Einstein-Straus black hole $t_{\rm ES}$.
Equations~\eqref{eq-ES.power-a} and \eqref{eq-ES.power-epsilon} give $\epsilon = (4 M)/(3 t_{\rm ret})$. 
Hence, equating $P_{\rm ES(4D)}$ to $-dM/dt_{\rm ret}$ in the left-hand side of eq.~\eqref{eq-ES.evapo-power.4D}, we can regard it as the evolution equation of mass $M$ as a function of the cosmological time $t$.
Up to the first correction term of order $O(\epsilon^{1/3})$, eq.~\eqref{eq-ES.evapo-power.4D} gives the semi-classical evolution equation of $M(t_{\rm ret})$ as
\eqb
 - \frac{d M}{dt} \simeq
 \frac{N}{30720 \pi\, M^2} \,
 \left[ 1 - 2 \left( \frac{4 M}{3 t} \right)^{1/3} \right] \, ,
\label{eq-ES.evapo-evolution}
\eqe
where we denote $t_{\rm ret}$ as $t$, representing the cosmological time of the evaporating black hole. 
Since the correction is negative, the emission is suppressed and the life time is prolonged.
Assuming that the correction term is small, we get the order estimate for the deviation of $t_{\rm ES}$ from the evaporation time of the Schwarzschild black hole $t_{\rm H}$ as
\eqb
 \frac{t_{\rm ES}}{t_{\rm H}} - 1 = O(M^{-2/3}) \, ,
\label{eq-ES.evapo-time}
\eqe
or 
\eqb
 t_{\rm ES} - t_{\rm H} = O(M^{7/3}) \, ,
\eqe
where $t_{\rm H}$ is given by neglecting the correction term in eq.~\eqref{eq-ES.evapo-evolution},
\eqb
 t_{\rm H} \simeq \frac{30720 \pi}{N} M^{3} \, .
\eqe
We can see from eq.~\eqref{eq-ES.evapo-time} that as the initial mass is larger, the evaporation time is better estimated by $t_{\rm H}$. 
This is reasonable since the Hubble parameter of the Einstein-de~Sitter universe becomes small as time proceeds and the cosmological effect on the evaporation becomes negligible. 
It should be noted that the cosmological correction on the evaporation time is relatively small even for a primordial black hole even if it was as large as the particle horizon unless its mass is of order the Planck mass. 
On the other hand, the deviation $(t_{\rm ES}-t_{\rm H})$ itself can be very large if the black hole
is very massive.

\section{Hawking Radiation from the Sultana-Dyer black hole}
\label{sec-SD}

\subsection{The Sultana-Dyer black hole}
\label{sec-SD.bg}

\begin{figure}[t]
 \begin{center}
 \includegraphics[height=65mm]{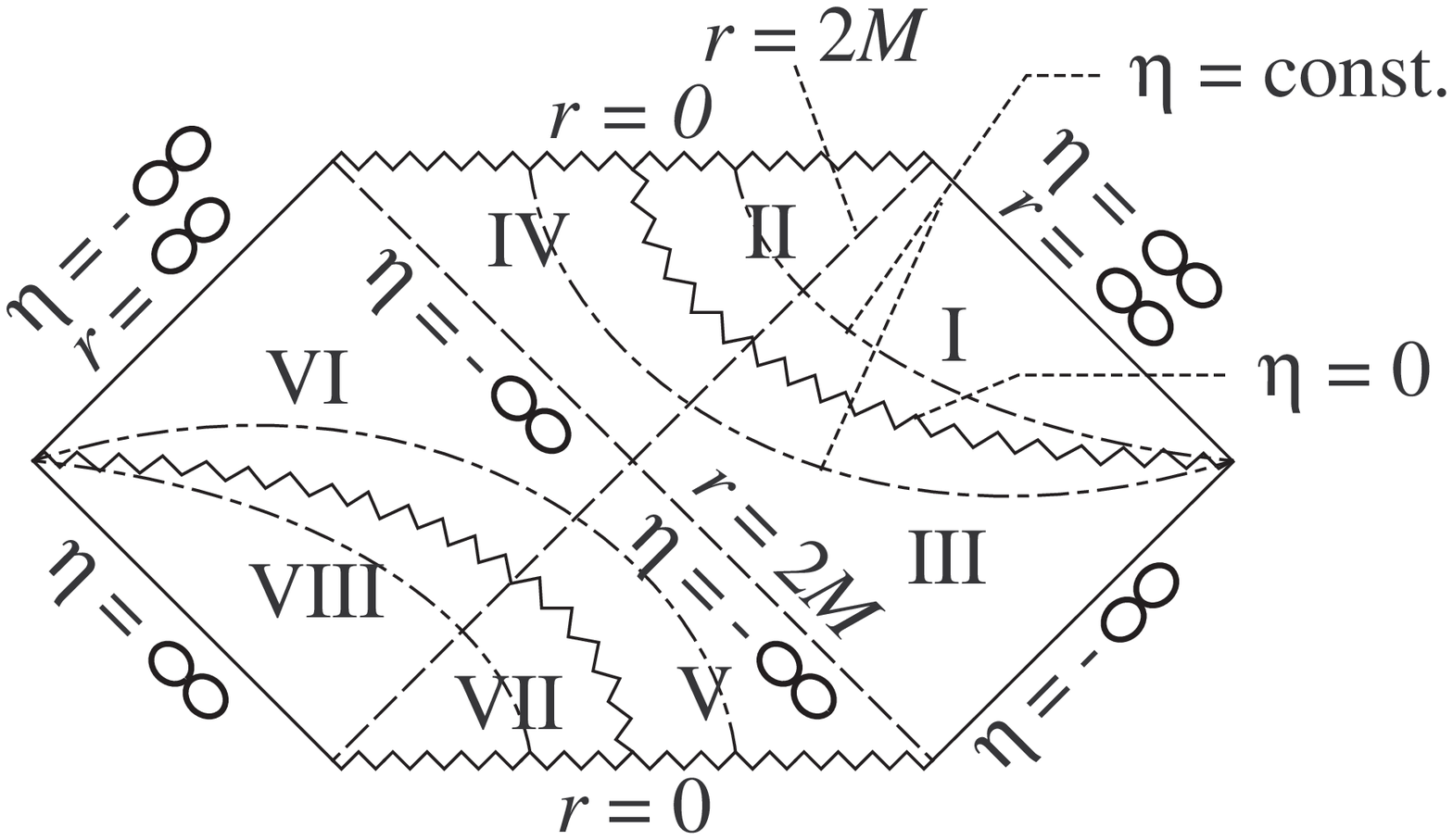}
 \includegraphics[height=50mm]{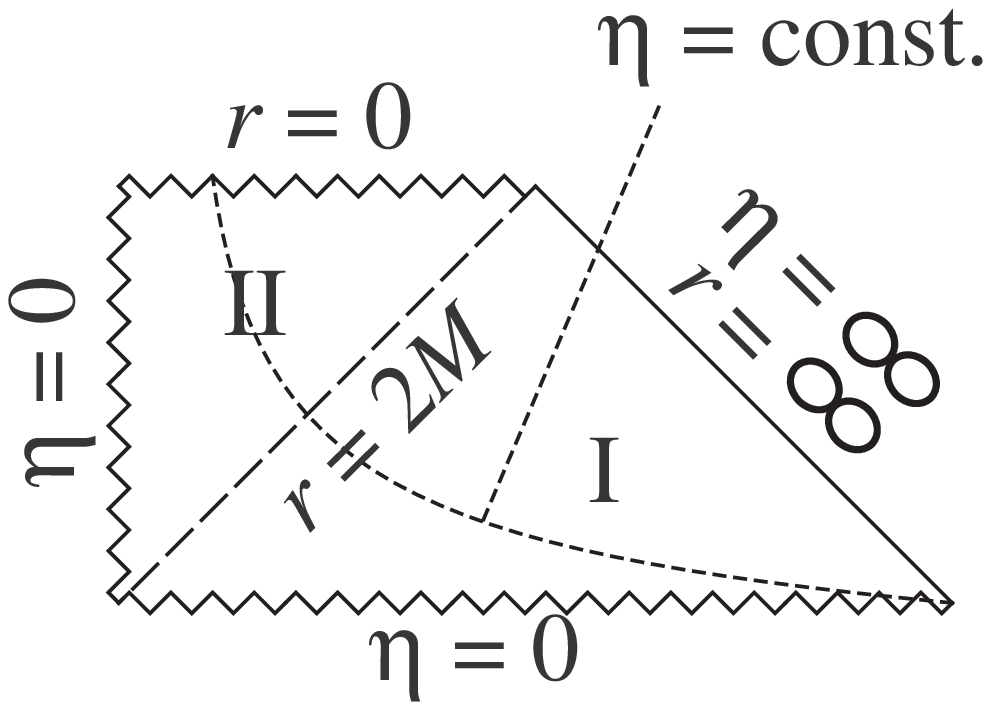}
 \end{center}
\caption{
The global structure of the Sultana-Dyer solution.
The zig-zag lines denote spacetime singularities, while the solid lines denote null infinity. 
So, the three pairs of regions, [I and II], [III, IV, V and VI] and [VII and VIII], respectively describe universes disconnected from each other.
Geodesics can reach the line of $\eta = -\infty$ and $r = 2M$ within finite affine length.
The spacetime given by Regions I and II is asymptotic to an expanding Friedmann region and represents a cosmological black hole.
}
\label{fig-3}
\end{figure}

The Sultana-Dyer black hole is obtained by the conformal transformation of the Schwarzschild black hole~\cite{ref-cosmo.bh.SD}. Its metric is given by
\eqb
 ds_{\rm SD}^2 =
 a(\eta)^2 \left[ - d\eta^2 + dr^2 + r^2 d\Omega^2
                  + \frac{2M}{r} (d\eta + dr)^2 \right] \, ,
\label{eq-SD.bg-SD}
\eqe
where $M$ is a positive constant, $a(\eta) = (\eta/\eta_{\ast})^2$ and $\eta_{\ast}$ is a constant. 
This spacetime is asymptotic to the Einstein-de~Sitter universe as $r \to \infty$. 
Here we consider the following coordinate transformation,
\eqb
 \eta \defeq t + 2 M \ln\left( \frac{r}{2 M} - 1 \right) \, .
\label{eq-SD.bg-coord.trans}
\eqe
This transforms the metric~\eqref{eq-SD.bg-SD} to the conformal Schwarzschild one,
\eqb
 ds_{\rm SD}^2 =
 a(t,r)^2\,\left[ - \left( 1 - \frac{2 M}{r} \right) dt^2
                  + \left( 1 - \frac{2 M}{r} \right)^{-1} dr^2 + r^2 d\Omega^2 \right] \, .
\eqe
$r=2M$ remains an event horizon because the conformal transformation preserves the causal structure.
The Penrose diagram of this spacetime is shown in fig.~\ref{fig-3}. 
There are curvature singularities at $\eta = 0$ and $r = 0$. 
The singularity at $\eta = 0$ is spacelike for $r>2M$, timelike for $r<2M$ and null for $r = 2M$. 
The central singularity at $r = 0$ is spacelike and surrounded by the event horizon. 
Hereafter we consider the spacetime given by regions I and II shown separately in the right panel of fig.~\ref{fig-3}.

This spacetime is conformally static, since there exists a conformal Killing vector $\xi = \partial_t$ which is the Killing vector on the Schwarzschild spacetime and satisfies the following relation
\eqb
 {\cal L}_{\xi}\, g_{\mu\nu} = \left({\cal L}_{\xi}\ln a^2 \right) \, g_{\mu\nu} \, ,
\label{eq-SD.bg-conf.Killing}
\eqe 
where $\partial_t = \partial_{\eta}$ due to the coordinate transformation \eqref{eq-SD.bg-coord.trans} and ${\cal L}_{\xi}\ln a^2 = 4/\eta$. 
The hypersurface at $r = 2M$ is the conformal Killing horizon which is a null hypersurface where $\xi$ becomes null. 
This coincides with the event horizon of the Sultana-Dyer black hole. 
The Misner-Sharp mass $m$ at an arbitrary spacetime point is given by
\eqb
 m(\eta,r) = M a - 2 M r a' + \frac{r^3 (a')^2}{2 a} \left( 1 + \frac{2M}{r} \right) \, ,
\eqe 
where $a' \defeq da/d\eta = 2\eta/\eta_{\ast}^2$. 
Then the Misner-Sharp mass at the event horizon is
\eqb
 m_{\rm EH} \defeq M a \left( 1 - \frac{8 M}{\eta} + \frac{32 M^2}{\eta^2} \right) \, .
\label{eq-SD.bg-mass}
\eqe
This means that the mass of the event horizon tends to increase in proportion to the scale factor $a\propto \eta^{2}$ as $\eta\to \infty$.

Substituting the metric \eqref{eq-SD.bg-SD} into the Einstein equation, we get
\eqb
 T^{\rm (SD)}_{\mu \nu} =
 \rho_{\rm m}\, u_{\mu}\, u_{\rm \nu} + \rho_{\rm r}\, k_{\mu}\, k_{\nu} \, ,
\eqe
where $u_{\mu}$ is a unit timelike vector, $k_{\mu}$ is a null vector normalised by $k^{\mu} u_{\mu} = -1$, $\rho_{\rm m}$ is the density of timelike dust and $\rho_{\rm r}$ is the density of null dust. The vectors $u^{\mu}$ and $k^{\mu}$ are given in $(\eta , r , \theta , \varphi)$ coordinates as
\seqb
\label{eq-SD.bg-velocities}
\eqab
 u^{\mu} &=&
  \left(
    \eta_{\ast}^2 \frac{r^2 + M ( 2r - \eta )}{r\,\eta^2\,\sqrt{r^2 + 2 M (r - \eta)}} \,,\,
  - \eta_{\ast}^2 \frac{M ( 2r - \eta )}{r\,\eta^2\,\sqrt{r^2 + 2 M (r - \eta)}} \,,\,
  0 \,,\, 0 \,\right) \, ,\\
 k^{\mu} &=& 
  \left(
    \eta_{\ast}^2 \frac{\sqrt{r^2 + 2 M (r - \eta)}}{r\,\eta^2} \,,\,
  - \eta_{\ast}^2 \frac{\sqrt{r^2 + 2 M (r - \eta)}}{r\,\eta^2} \,,\,
  0 \,,\, 0 \,\right) \, .
\eqae
\seqe
The densities of dusts are given by
\seqb
\label{eq-SD.bg-densities}
\eqab
  \rho_{\rm m} &=&
   \eta_{\ast}^4 \frac{12 \left[ r^2 + 2 M (r - \eta) \right]}{8 \pi\, r^2\, \eta^6} \, ,\\
  \rho_{\rm r} &=&
   \eta_{\ast}^4 \frac{4 M \left[ 4 r^2 + 3 M (2 r - \eta) \right]}
                        {8 \pi\, r^2\, \eta^5 \left[ r^2 + 2 M (r - \eta) \right]} \, .
\eqae
\seqe
These densities imply that energy conditions are satisfied only when $\eta < r (r + 2 M)/2 M$ where $\rho_{\rm m} > 0$ and $\rho_{\rm r} > 0$. 
Furthermore the velocities of dusts \eqref{eq-SD.bg-velocities} denote that the null dusts in the region $\eta < r (r + 2 M)/2 M$ fall towards the black hole, and also that the timelike dusts in the region $\eta < r (r + 2 M)/2 M$ and $\eta < 2 r$ do so. 
We can describe a physical picture that the accretion of timelike and null dusts increases the mass of black hole as shown in eq.~\eqref{eq-SD.bg-mass}. 
For $\eta > r(r + 2M)/2M$, the source matter fields of the Einstein equation get unphysical. 
However, the Sultana-Dyer metric is featured with the global structure of a cosmological black hole as seen in Fig.~\ref{fig-3} and also with the conformally static nature which makes the physical interpretation and calculation of quantum stress-energy tensor most tractable. 
Hence, we adopt it as a not only workable but also physically interesting model for a cosmological black hole with significant mass accretion.

One may consider the future outer trapping horizon as a local definition of a black hole horizon~\cite{ref-hayward}. 
The trapping horizon is given by $2 m(\eta,r) = R(\eta)$, where $R = r a$ is an areal radius. 
Thus the trapping horizon in this spacetime is obtained by the following algebraic equation:
\eqab
 1 = \frac{2M}{r} - \frac{8M}{\eta} + \frac{4r^2}{\eta^{2}}\left(1 + \frac{2M}{r}\right) \, .
\eqae 
This has two roots, $r=r_{1}$ and $r=r_{2}$ ($r_{1}<r_{2}$), where
\seqb
\eqab
 r_{1} &\defeq& - M + \frac{-\eta + \sqrt{\eta^{2} + 24 M \eta + 16 M^{2}}}{4}, \\
 r_{2} &\defeq& \frac{\eta}{2}.
\eqae
\seqe
Regions $0 < r < r_{1}$, $r_{1} < r < r _{2}$ and $r_{2} < r$ are future trapped, untrapped and past trapped, respectively.

\subsection{Power of Hawking radiation from the Sultana-Dyer black hole}
\label{sec-SD.power}

We introduce a matter field which describes quantum radiation from the black hole. 
Let $\phi$ be a massless scalar field with conformal coupling, satisfying $( \Box - {\cal R}/6 ) \phi = 0$, where ${\cal R}$ is the Ricci scalar, for which we will see that the renormalised stress-energy tensor 
$\tmn{\mu \nu}$ can be expressed in terms of $\tmn{\mu \nu}_{\rm Sch}$ for the Schwarzschild spacetime. 
Appendix summarises how to derive $\tmn{\mu \nu}$. 
We use eq.~\eqref{eq-app.four-stress} given in Appendix. 
For the Sultana-Dyer spacetime, the metric $\tg_{\mu \nu}$ and the tensors $\widetilde{X}_{\mu \nu}$ and $\widetilde{Y}_{\mu \nu}$ in eq.~\eqref{eq-app.four-stress} are all for the Schwarzschild spacetime, and $\ttmn{\mu \nu}$ is $\tmn{\mu \nu}_{\rm Sch}$. 
Obviously, $\widetilde{X}_{\mu \nu} = \widetilde{Y}_{\mu \nu} = 0$ since $\widetilde{R}_{\mu \nu} = 0$. Then eq.~\eqref{eq-app.four-stress} becomes
\eqb
 \tmn{\mu \nu}
 = \frac{1}{a^2}\, \tmn{\mu \nu}_{\rm Sch}
   - \frac{1}{2880 \pi^2}\,\left( \frac{1}{6} X_{\mu \nu} - Y_{\mu \nu} \right) \, ,
\label{eq-SD.power-stress}
\eqe
where $X_{\mu \nu}$ and $Y_{\mu \nu}$ are obtained by substituting the metric \eqref{eq-SD.bg-SD} into eqs.~\eqref{eq-app.four-XY}. 
The first term in $\tmn{\mu \nu}$ expresses purely the Hawking radiation from black hole, and the second term includes cosmological particle creation. 
Hereafter we consider an observer at $r =$ const, i.e., we calculate $\tmn{\mu \nu}$ in $(\eta,r,\theta,\varphi)$ coordinates of eq.~\eqref{eq-SD.bg-SD}, which implies that the vacuum state $\vac$ in $\tmn{\mu \nu}$ is defined with respect to the mode function of $\phi$ in the same coordinates.

For the observer distant from the black hole, the observed energy flux $F_{\rm obs}$ is given by
\eqb
 F_{\rm obs} \defeq
   \langle T_{(\eta)}^{(r)} \rangle
 = - \frac{1}{a^4} \tmn{\eta r}_{\rm Sch}
   + \frac{1}{a^2} \, \frac{1}{2880 \pi^2}\,
     \left( \frac{1}{6} X_{\eta r} - Y_{\eta r} \right) \, ,
\label{eq-SD.power-flux.obs.1}
\eqe
where $\langle T_{(\eta)}^{(r)} \rangle$ is a tetrad component evaluated in the Sultana-Dyer spacetime, and eq.~\eqref{eq-SD.power-stress} and $T_{(\eta)}^{(r)} = - a^{-2} T_{\eta r}$ at the distant region are used in the second equality. 
The distant observer is comoving with the timelike dust as seen in eq.~\eqref{eq-SD.bg-velocities}. 
From eqs.~\eqref{eq-app.four-XY}, the second term in the right-hand side of eq.~\eqref{eq-SD.power-flux.obs.1} becomes
\eqb
\begin{split}
 \frac{1}{2880 \pi^2} & \left( \frac{1}{6} X_{\eta r} - Y_{\eta r} \right) \\
 =&\, - \frac{M^3\, \eta_{\ast}^8}{30 \pi^2\, \eta^9\, r^6}
 + \frac{M^2\, \eta_{\ast}^4\, \left[- 9 M \eta^8 + (21 M + \eta) \eta_{\ast}^4 \eta^4
                                       + 12 M \eta_{\ast}^8 \right]}
        {90 \pi^2\, \eta^{14}\, r^5} \\
 &-\, \frac{M^2\, \eta_{\ast}^4\, \left[- 88 M \eta^8 + (28 M - \eta) \eta_{\ast}^4 \eta^4
                                       + 160 M \eta_{\ast}^8 \right]}
        {120 \pi^2\, \eta^{15}\, r^4} \\
 &-\, \frac{M\, \eta_{\ast}^4\,
          \left[ 2 M (65 M - 37 \eta) \eta^8
               + (-240 M^2 - 64 M \eta + 11 \eta^2) \eta_{\ast}^4 \eta^4
               - 120 M (5 M - \eta) \eta_{\ast}^8 \right]}
        {180 \pi^2\, \eta^{16}\, r^3} \\
 &+\, \frac{M\, \eta_{\ast}^4\,
          \left[- 4 M (65 M - 27 \eta) \eta^8 + 5 (96 M + 11 \eta) \eta_{\ast}^4 \eta^4
                + 1200 M \eta_{\ast}^8 \right]}
        {360 \pi^2\, \eta^{16}\, r^2} \\
 &-\, \frac{M\, \eta_{\ast}^4\,
          \left[ 13 \eta^8 - 24 \eta_{\ast}^4 \eta^4 - 60 \eta_{\ast}^8 \right]}
        {72 \pi^2\, \eta^{16}\, r} \, .
\end{split}
\eqe
This falls off very rapidly for a distant observer, and hence we get $F_{\rm obs} = - a^{-4}\,\tmn{\eta r}_{\rm Sch} $ for the distant observer. 
Here the coordinate transformation~\eqref{eq-SD.bg-coord.trans} gives $\tmn{\eta r}_{\rm Sch} = \tmn{t r}_{\rm Sch} + (1 - r/2M)^{-1} \tmn{t t}_{\rm Sch}$. Hence we obtain
\eqb
 F_{\rm obs}
 = - \frac{1}{a^4} \, \tmn{t r}_{\rm Sch}
 = \frac{1}{a^4}\, \langle T_{(r)}^{(t)} \rangle_{\rm Sch} \, ,
\label{eq-SD.power-flux.obs.2}
\eqe
where $\langle T_{(r)}^{(t)} \rangle_{\rm Sch}$ is a tetrad component evaluated in the Schwarzschild spacetime. 
This $F_{\rm obs}$ is the flux (energy flow per unit time and unit area) detected by the distant comoving observer.

It should be pointed out that the flux $F_{\rm obs}$ contains only the Hawking radiation from black hole but no cosmological particle creation, since $F_{\rm obs}$ is proportional only to $\tmn{t r}_{\rm Sch}$ as shown in eq.~\eqref{eq-SD.power-flux.obs.2}. 
This does not imply the absence of cosmological particle creation at our distant observer. 
The energy density $\sigma$ of the quantum field $\phi$ in the distant region indicates the cosmological particle creation at the distant observer. 
We get from eqs.~\eqref{eq-SD.power-stress} and \eqref{eq-app.four-XY},
\eqb
 \begin{split}
  \sigma
   &\defeq \langle T_{(\eta)}^{(\eta)} \rangle
         = - \frac{1}{a^2}\, \tmn{\eta \eta} \\
   &= - \frac{1}{a^4}\, \tmn{\eta \eta}_{\rm Sch}
      + \frac{59 \eta_{\ast}^8 \eta^8 + 70 \eta_{\ast}^{12} \eta^4
              - 100 \eta_{\ast}^{16}}
             {240 \pi^2 \eta^{20}} \, ,
 \end{split}
\eqe
for the distant observer.
The second term does not include $M$ and expresses purely the cosmological particle creation in the distant region. 
However this raises no energy flow as shown in eq.~\eqref{eq-SD.power-flux.obs.2}.

Hence $F_{\rm obs}$ in eq.~\eqref{eq-SD.power-flux.obs.2} is the flux of the Hawking radiation from Sultana-Dyer black hole.
Then the intrinsic power of the Hawking radiation $P_{\rm SD}$ should be given by
\eqb
 P_{\rm SD} \defeq
   \left. 4 \pi ( r a \right)^2\, F_{\rm obs}
 = \frac{1}{a^2} \,
   \left[ 4 \pi r^2
          \langle T_{(r)}^{(t)} \rangle_{\rm Sch} \right] \, ,
\eqe
where the right-hand side should be evaluated for the distant observer. 
Here note that the factor $4 \pi r^2 \langle T_{(r)}^{(t)} \rangle_{\rm Sch} $ is the observed power of the Hawking radiation in the Schwarzschild spacetime, and it should equal $P_{\rm H(4D)}$ given in eq.~\eqref{eq-ES.evapo-power.4D.flat} if the geometrical optics approximation is valid. 
Therefore, under this approximation, we get
\eqb
 P_{\rm SD} = \frac{1}{a^2}\, \frac{N\,\kappa^2}{1920 \pi} \, ,
\label{eq-SD.power-power}
\eqe
where $\kappa = 1/(4M)$ and $N$ is given by eq.~\eqref{eq-ES.evapo-N}.

The geometrical optics approximation gets very good for late times, and the mass $m_{\rm EH}(\eta)$ in eq.~\eqref{eq-SD.bg-mass} becomes $m_{\rm EH} \to M a$ as $\eta \to \infty$. 
Hence, by comparing our result $P_{\rm SD}$ with the Schwarzschild one $P_{\rm H(4D)}$, it is suggested that the effective temperature $T_{\rm eff}$ of the Sultana-Dyer black hole at late times is given by
\eqb
 T_{\rm eff} = \frac{1}{8 \pi\, M a} = \frac{\kappa}{2 \pi\, a} \, .
\label{eq-SD.power-temperature}
\eqe
So, both the intrinsic power and temperature of the radiation from the Sultana-Dyer black hole are the same as those of the Hawking radiation from the Schwarzschild black hole of which mass is the momentary mass of the growing event horizon.
This temperature $T_{\rm eff}$ and the intrinsic power $P_{\rm SD}$ decrease as time proceeds. 
This result is reasonable since the Sultana-Dyer black hole describes significant mass accretion. 
This black hole does not lose but gain mass due to the accretion of timelike and null dusts, and can be regarded as an object in quasi-equilibrium with temperature $T_{\rm eff}$.

\subsection{Conformal dynamics at infinity and temperature of Sultana-Dyer black hole}
\label{sec-SD.temp}

A stationary spacetime is defined by a timelike Killing vector $\zeta$ satisfying ${\cal L}_{\zeta}\, g_{\mu\nu} = 0$.
The Killing horizon is a null hypersurface where $\zeta$ becomes null, and the surface gravity $\kappa$ is defined by $\zeta^{\alpha}\nabla_{\alpha} \zeta^{\mu} = \kappa\,\zeta^{\mu}$ at the Killing horizon. 
The value of $\kappa$ changes according to the normalisation of $\zeta$ by definition. 
For asymptotically flat stationary black hole spacetimes, the Killing horizon coincides with the event horizon. 
The surface gravity $\kappa$ of the stationary black hole is constant everywhere on the event horizon. 
This is the zeroth law of black hole thermodynamics~\cite{ref-bht,ref-hr} which states that a unique temperature can be assigned to the stationary black hole. 
Then the thermal spectrum of Hawking radiation from the stationary black hole, which is a quantum phenomenon, determines the value of the temperature to be $\kappa/2 \pi$ under the normalisation of Killing vector as $\zeta^{\mu}\zeta_{\mu} \to -1$ at null and spatial infinities~\cite{ref-hr}.

Several generalisations of the zeroth law have already been discussed for general {\it conformal stationary black hole spacetimes} whose metric $g_{\mu\nu}$ is given by $g_{\mu\nu} = \Omega^2\, \tilde{g}_{\mu\nu}$ where $\Omega^2$ is the conformal factor and $\tilde{g}_{\mu\nu}$ is the metric of asymptotically flat stationary black hole~\cite{ref-zeroth.DH, ref-zeroth.SD, ref-zeroth.JK}. 
A natural generalisation of the surface gravity $\kappa_{\rm DH}$ can be introduced by the following relation at the conformal Killing horizon~\cite{ref-zeroth.DH, ref-zeroth.SD},
\eqb
 \xi^{\alpha}\nabla_{\alpha} \xi^{\mu} = \kappa_{\rm DH}\, \xi^\mu \, ,
\label{eq-SD.temp-sg.DH}
\eqe 
where $\xi$ is a conformal Killing vector satisfying ${\cal L}_{\xi}\,g_{\mu \nu} = ( {\cal L}_{\xi} \ln\Omega^2 )\, g_{\mu \nu}$, and the conformal Killing horizon is the hypersurface where $\xi$ becomes null. 
Under the conditions $\Omega \to 1$ (or constant) and $\xi^{\mu} \xi_{\nu} \to -1$ at null infinity, Sultana and Dyer conjectured that the temperature of conformal stationary black holes $T_{\rm SD}$ is given by~\cite{ref-zeroth.SD} 
\eqb
 T_{\rm SD} \defeq
 \frac{1}{2 \pi}\, \left( \kappa_{\rm DH} - {\cal L}_{\xi} \ln\Omega^2 \right) \, .
\eqe
$T_{\rm SD}$ is constant everywhere on the conformal Killing horizon, while $\kappa_{\rm DH}$ is not. 
On the other hand, Jacobson and Kang independently introduced a generalised surface gravity $\kappa_{\rm JK}$ as~\cite{ref-zeroth.JK},
\eqb
 \nabla_{\mu} \left( \xi^{\alpha} \xi_{\alpha} \right) = - 2 \kappa_{\rm JK}\, \xi_{\mu} \, .
\label{eq-SD.temp-sg.JK}
\eqe
$\kappa_{\rm JK}$ is invariant under the conformal transformation, while $\kappa_{\rm DH}$ is not. 
Then under the conditions $\Omega \to 1$ and $\xi^{\mu} \xi_{\nu} \to -1$ at null infinity, they conjectured that the temperature of conformal stationary black holes $T_{\rm JK}$ is given by
\eqb
 T_{\rm JK} \defeq \frac{\kappa_{\rm JK}}{2 \pi} \, .
\eqe
In fact, it can be shown that the relation $\kappa_{\rm JK} = \kappa_{\rm DH} - {\cal L}_{\xi} \ln\Omega^2$ holds~\cite{ref-zeroth.JK}, and hence $T_{\rm SD} = T_{\rm JK} \defeqr T_{\rm JKSD}$. 
Thus, although Sultana-Dyer~\cite{ref-zeroth.SD} and Jacobson-Kang~\cite{ref-zeroth.JK} considered independently the surface gravity and temperature of conformal stationary black holes, they reached the same conjecture.

For the Sultana-Dyer black hole, the conjectured temperature $T_{\rm JKSD}$ has already been calculated~\cite{ref-cosmo.bh.SD}. 
The conformal Killing vector is $\xi = \partial_{\eta}$ and the conformal Killing horizon coincides with the event horizon $r = 2M$. 
The norm of $\xi$ is $\xi^{\mu} \xi_{\mu} \to -a^2 \neq -1$ at null infinity, not satisfying the unit norm condition for $T_{\rm JKSD}$.
Sultana and Dyer still assumed in~\cite{ref-cosmo.bh.SD} the temperature $T_{\rm JKSD}$ should be assigned to the Sultana-Dyer black hole. 
Then, substituting $\xi$ into eq.~\eqref{eq-SD.temp-sg.DH} or \eqref{eq-SD.temp-sg.JK}, they obtained $\kappa_{\rm JK} = \kappa_{\rm DH} - 4/\eta = 1/(4 M)$ and 
\eqb
 T_{\rm JKSD} = \frac{1}{8 \pi M} \, ,
\eqe 
which is equal to the Hawking temperature of the Schwarzschild black hole of mass $M$, but not to our effective temperature $T_{\rm eff}$ in eq.~\eqref{eq-SD.power-temperature}. 
However, the temperature of black hole should be given based on the spectrum and/or the power of the Hawking radiation. 
Here, we propose that the physically reasonable temperature of black holes which are conformal stationary and asymptotically dynamical is not $T_{\rm JKSD}$ but 
\eqb
 T_{\rm eff} \defeq \frac{T_{\rm JKSD}}{\Omega} \, . 
\label{eq-SD.temp-temp}
\eqe
So the effective temperature depends on space and time through the conformal factor.
This might be understood in an analogy with Tolman's law for thermal equilibrium in the presence of a gravitational field~\cite{ref-tolman}.

\section{Summary and discussions}
\label{sec-sd}

We have calculated the intrinsic power of the Hawking radiation from cosmological black holes for two cases, no mass accretion and significant mass accretion. 

For no mass accretion case, we have considered the Einstein-Straus black hole. 
Our result $P_{\rm ES(4D)}$ in eq.~\eqref{eq-ES.evapo-power.4D} indicates $P_{\rm ES(4D)} < P_{\rm H(4D)}$, i.e., the black hole evaporation is suppressed by the cosmological expansion.
The ratio $P_{\rm ES(4D)}/P_{\rm H(4D)} \, (<1)$ is given in terms of $\epsilon^{1/3}$ where $\epsilon$ is the ratio in size of the black hole to the cosmological horizon. 
The first correction term is $O(\epsilon^{1/3})$ and therefore currently as small as $10^{-5}(M/10^{6}M_{\odot})^{1/3} (t/14\, \mbox{Gyr})^{-1/3}$, but could be significant for the formation epoch of primordial black holes. 
The evaporation time is essentially the same as that of the Schwarzschild black hole as long as its mass is greater than the Planck mass. 
Furthermore, by comparing the functional form of $P_{\rm ES(4D)}$ with that of $P_{\rm H(4D)}$ in eq.~\eqref{eq-ES.evapo-power.4D.flat}, we can see that the 
Einstein-Straus black hole has not settled down to thermal equilibrium 
in a finite cosmological time.

For the significant mass accretion case, we have considered the Sultana-Dyer black hole. 
This has very different properties. 
Our result $P_{\rm SD}$ in eq.~\eqref{eq-SD.power-power} indicates that the Sultana-Dyer black hole does not evaporate away.
Furthermore the Sultana-Dyer black hole can be regarded as an object in quasi-equilibrium, since the effective temperature $T_{\rm eff}$ can be assigned as eq.~\eqref{eq-SD.power-temperature}. 
The intrinsic power $P_{\rm SD}$ of the Hawking radiation is consistent with the Stefan-Boltzmann law for a black body with temperature $T_{\rm eff}$. 
We propose a new definition \eqref{eq-SD.temp-temp} for the temperature for general conformal stationary black holes.

Finally we try to interpret $P_{\rm ES(4D)}$ for the Einstein-Straus black hole in an analogy with quantum radiation of a slowly moving mirror in a flat spacetime. 
For simplicity, we consider a moving mirror $x=x(t)$ in the two dimensional Minkowski spacetime with the Cartesian coordinates~$(t,x)$. 
Then it is well known that the moving mirror emits quantum radiation of a massless scalar field $\phi$.
When an observer at rest is in the region $x > x(t)$, the observed power $P_{\rm mir}$ of quantum radiation from the mirror is given by~\cite{ref-qft,ref-mirror}
\eqb
 P_{\rm mir} \defeq \langle T_{(t)}^{(x)} \rangle_{\rm mir}
 = - \frac{1}{12\pi}\,\frac{\sqrt{1 - v^2}}{(1 - v)^2}\,\frac{d \alpha_{\rm mir}}{dt_{\rm ret}}
 \, ,
\eqe
where $\langle T_{(t)}^{(x)} \rangle_{\rm mir}$ is a tetrad component, $t_{\rm ret}$ is the retarded time when the observed particle of $\phi$ was emitted from the mirror, $v = dx(t)/dt |_{t_{\rm ret}} \defeq \dot{x}(t_{\rm ret})$ is positive when the mirror is approaching towards the rest observer, and $\alpha_{\rm mir}$ is the proper acceleration of the mirror. 
In the Minkowski spacetime, the observed power $P_{\rm mir}$ is equal to the intrinsic power of quantum radiation from the mirror. 
Here we consider the case that the mirror moves slowly, i.e., $|v| \ll 1$. 
Then the power $P_{\rm mir}$ is given as
\eqb
 P_{\rm mir} \simeq
 - \frac{\dddot{x}(t_{\rm ret})}{12 \pi} \,
 \left( 1 + 2\, v \right)+O(v^{2}) \, .
\label{eq-sd-power.mirror}
\eqe
Note that the kinematic effect comes in the radiation power in the form of $(1+2v)$ in the lowest order.
On the other hand for the Einstein-Straus black hole, the relative ``velocity'' $v_{\rm ES}$ of the junction surface $\Sigma$ to the black hole at the retarded time may be written as 
\eqb
 v_{\rm ES} = - r_{\rm \Sigma}\, \dot{a}_{\rm ret} \, ,
\eqe
where $a_{\rm ret} \defeq a(t_{\rm ret})$, and the minus sign means the increase of the relative distance. 
Equation~\eqref{eq-ES.bg-jc.a} with $k=0$ gives $\dot{a}_{\rm ret}^3 = (2 M H_{\rm ret})/r_{\rm \Sigma}^3 = \epsilon/r_{\rm \Sigma}^3$, i.e., $v_{\rm ES} = - \epsilon^{1/3}$. 
Therefore, the kinematic correction factor in eq.~\eqref{eq-sd-power.mirror} coincides with the cosmological correction factor in eq.~\eqref{eq-ES.evapo-power.4D} up to this order.
Hence, we can interpret the correction factor $(1-2\,\epsilon^{1/3})$ in $P_{\rm ES(4D)}$ is some kinematic effect from the cosmological expansion in an analogy with radiation from a moving mirror.

\acknowledgments

TH and HM were respectively supported by the Grant-in-Aid for Scientific Research Fund of the Ministry of Education, Culture, Sports, Science and Technology, Japan (Young Scientists (B) 18740144 and 18740162). 
HM was also supported by the Grant Nos. 1071125 from FONDECYT (Chile).
CECS is funded in part by an institutional grant from Millennium Science Initiative, Chile, and the generous support to CECS from Empresas CMPC is gratefully acknowledged.

\appendix*
\section{Vacuum expectation value of quantum stress-energy tensor}

\subsection{Two dimensional case}

It has already been recognised for a few decades that many different methods of renormalisation give equivalent results (see for example, chapters 6 and 7 in~\cite{ref-qft}). 
We consider a minimally coupled massless scalar field $\phi$, whose stress-energy tensor is given by
\eqb
 T_{\mu \nu} =
 \phi_{,\mu} \phi_{,\nu} - \frac{1}{2} g_{\mu \nu} \phi_{,\alpha} \phi^{,\alpha} \, .
\label{eq-app.two-c.stress}
\eqe
The background spacetime is described in double null coordinates $(u,v)$ as
\eqb
 ds^2 = -  D(u,v) \, du\, dv \, .
\eqe

The field $\phi$ satisfies the Klein-Gordon equation $\Box\phi = 0$. 
When a coordinate system (not necessarily null) is specified to describe the differential operator $\Box$, we can find a complete orthonormal set $\{f_{\omega}\}$ for arbitrary solutions of $\Box\phi = 0$, where $\omega$ denotes the frequency of the mode function. 
The positive frequency mode is the mode function $f_{\omega}$ which is constructed to satisfy the conditions, $\omega > 0$~, $\left( f_{\omega} , f_{\omega'} \right) = \delta(\omega - \omega')$~, $(f_{\omega} , f_{\omega'}^{\ast}) = 0$ and $\left( f_{\omega}^{\ast} , f_{\omega'}^{\ast} \right) = - \delta(\omega - \omega')$, where $(f,g)$ is the inner-product defined from the Noether charge of time translation of $\phi$ and $f_{\omega}^{\ast}$ is complex conjugate to $f_{\omega}$, called the negative frequency mode. 
In two dimensional spacetimes, the positive frequency modes can be decomposed with respect to the direction of propagation. 
In the double null coordinates, they are $f_{\omega}(u) = \exp(-i\,\omega\,u)/\sqrt{4 \pi \omega}$ and $f_{\omega}(v) = \exp(-i\,\omega\,v)/\sqrt{4 \pi \omega}$. 
Then, the quantum operator $\phi$ is expanded by the complete orthonormal set of the positive and negative frequency modes as
\eqab
 \phi(u,v) = \int_0^{\infty} d\omega\,
 \left[\, a_{\omega}\, f_{\omega}(u)
        + a_{\omega}^{\dag}\, f_{\omega}^{\ast}(u)
        +  b_{\omega}\, f_{\omega}(v)
        + b_{\omega}^{\dag}\, f_{\omega}^{\ast}(v) \,\right] \, .
\label{eq-app.two-expand}
\eqae
The canonical quantisation presumes the simultaneous commutation relation between $\phi$ and its conjugate momentum, so that $\{a_{\omega}\}$ and $\{b_{\omega}\}$ are harmonic operators satisfying the commutation relations; 
$[ a_{\omega} , a_{\omega'}^{\dag} ] = \delta(\omega - \omega')$ and $ [ b_{\omega} , b_{\omega'}^{\dag} ] = \delta(\omega - \omega')$ and all others vanish. 
They define the Fock space of quantum states and give particle interpretation. 
The vacuum state $\vac$ is defined as a quantum state satisfying $a_{\omega} \vac = b_{\omega} \vac = 0$ for all $\omega$.

If we choose different coordinates $(\bar{u},\bar{v})$, a natural orthonormal set of mode functions is $\{\bar{f}_{\omega}\}$, where $\bar{f}_{\omega}(\bar{u}) = \exp(-i\,\omega\,\bar{u})/\sqrt{4 \pi \omega}$ and $\bar{f}_{\omega}(\bar{v}) = \exp(-i\,\omega\,\bar{v})/\sqrt{4 \pi \omega}$. 
Then the expansion \eqref{eq-app.two-expand} gives different harmonic operators $\{\bar{a}_{\omega}\}$ and $\{\bar{b}_{\omega}\}$. These harmonic operators define another vacuum state $\bvac \,(\,\neq \vac \,)$ if there arises the mixing of positive and negative frequency modes $\left( f_{\omega} , \bar{f}_{\omega'}^{\ast} \right) \not\equiv 0$ between the two coordinate systems. 
Thus, even if a quantum state is initially set to be a vacuum state, this does not remain vacuum but corresponds to an excited state associated with the coordinate system natural to an observer at the final time if the mixing of positive and negative modes arises. 
This will be interpreted as quantum particle creation in curved spacetimes.

The quantum expectation value of the stress-energy tensor $\bvacbra T_{\mu \nu} \bvac$ is 
calculated by substituting the quantum operator \eqref{eq-app.two-expand} (after replacing $a_{\omega}$ and $b_{\omega}$ with $\bar{a}_{\omega}$ and $\bar{b}_{\omega}$) into the stress-energy tensor \eqref{eq-app.two-c.stress}. 
However, $\bvacbra T_{\mu \nu} \bvac$ diverges even for flat background cases. 
Therefore, we need to renormalise the stress-energy tensor. 
We do not get into the details of the regularisation method but only quote the result~\cite{ref-expect.stress.2D},
\eqab
 \tmn{\bm \bn} = \theta_{\bm \bn} + \frac{{\cal R}}{48\, \pi}\, g_{\bm \bn} \, ,
\label{eq-app.two-stress.1}
\eqae
where $\tmn{\bm \bn}$ is the renormalised expectation value of $\bvacbra T_{\bm \bn} \bvac$, ${\cal R}$ is the Ricci scalar of the background spacetime, and $\theta_{\bm \bn}$ is a symmetric tensor whose components in the coordinate system $(\bu,\bv)$ on which the vacuum $\bvac$ is defined is given by
\seqb
\label{eq-app.two-stress.2}
\eqab
 \theta_{\bu \bu} &\defeq&
   -\frac{1}{24 \pi} \left[ \frac{3}{2} \left(\frac{D_{,\bu}}{D}\right)^2
                          - \frac{D_{,\bu \bu}}{D} \right] , \\
 \theta_{\bv \bv} &\defeq&
   -\frac{1}{24 \pi} \left[ \frac{3}{2} \left(\frac{D_{,\bv}}{D}\right)^2
                          - \frac{D_{,\bv \bv}}{D} \right] , \\
 \theta_{\bu \bv} &=& \theta_{\bv \bu} \equiv 0 ,
\eqae
\seqe
where $D(\bu,\bv) = - 2\, g_{\bu \bv}$. 
The renormalised expectation value $\tmn{\mu \nu}$ of $\bvacbra T_{\mu \nu} \bvac$ in the other coordinates $(u,v)$ is calculated from the above components through the usual coordinate transformation for tensor components,
\eqab
 \tmn{\mu \nu} =
  \frac{\partial x^{\bm}}{\partial x^{\mu}}\,
  \frac{\partial x^{\bn}}{\partial x^{\nu}}\, \tmn{\bm \bn} \, .
\eqae

\subsection{Four dimensional case}

The renormalised vacuum expectation value of stress-energy tensor $\tmn{\mu \nu}$ of some matter field in four dimensions may also be calculated along with the canonical quantisation formalism as shown for the two dimensional case in previous section. 
However the path integral quantisation formalism is more convenient to summarise $\tmn{\mu \nu}$ on a four dimensional conformal spacetime. 
The effective action $W$ of a quantum matter field $\phi$ on a spacetime of metric $g_{\mu \nu}$ gives the vacuum expectation value of quantum stress-energy tensor.
$W$ can be evaluated by the path integral method and the vacuum state $\vac$ is specified by the Green function of $\phi$ used in evaluating the path integral. 
However the precise path integral form of $W$ is not important here. 
$W$ is decomposed into two parts as $W = W_{\rm ren} + W_{\rm div}$, where $W_{\rm ren}$ is the renormalised part and $W_{\rm div}$ is the divergent part. 
The functional differentiation of $W_{\rm ren}$ gives the renormalised vacuum expectation value $\tmn{\mu \nu}$,
\eqb
 \tmn{\mu \nu} = \frac{2}{\sqrt{-g}}\, \frac{\delta W_{\rm ren}}{\delta g^{\mu \nu}} \, .
\label{eq-app.four-stress.1}
\eqe

We consider the case that the metric $g_{\mu \nu}$ is conformal to the other one as
\eqb
 g_{\mu \nu} = \Omega^2\, \tg_{\mu \nu} \, ,
\label{eq-app.four-conf.metric}
\eqe
and the matter field $\phi$ is a conformally coupled massless scalar field satisfying $(\Box - {\cal R}/6) \phi = 0$. On the other hand, we get by definition of functional differentiation,
\eqb
 W_{\rm ren} - \tW_{\rm ren} =
 \int \frac{\delta W_{\rm ren}}{\delta g^{\alpha \beta}}\,\delta g^{\alpha \beta}\, d^4x \, ,
\label{eq-app.four-action}
\eqe
where $\tW_{\rm ren}$ is the renormalised effective action obtained from $W_{\rm ren}$ with replacing $g_{\mu \nu}$ by $\tg_{\mu \nu}$. 
Then considering the functional differentiation only by the conformal transformation, $\delta g^{\mu \nu} = - 2 g^{\mu \nu}\,\Omega^{-1}\, \delta\Omega$, the effective action is expressed as
\eqb
 W_{\rm ren} =
 \tW_{\rm ren} - \int g^{\alpha \beta}\tmn{\alpha \beta}\,
                 \frac{\delta\Omega}{\Omega}\, \sqrt{-g}\, d^4x \, .
\eqe
Substituting this into eq.~\eqref{eq-app.four-stress.1}, we get
\eqb
 \tmn{\mu \nu} =
   \frac{1}{\Omega^{2}} \ttmn{\mu \nu}
 - \frac{2}{\sqrt{-g}} \, \frac{\delta}{\delta g^{\mu \nu}}\,
   \int g^{\alpha \beta}\tmn{\alpha \beta}\, \frac{\delta\Omega}{\Omega}\, \sqrt{-g}\, d^4x \, ,
\label{eq-app.four-stress.2}
\eqe
where $\widetilde{\delta}_{\mu}^{\nu} = \tg_{\mu \alpha} \tg^{\alpha \nu}$, $g_{\mu \sigma}\tg^{\sigma \alpha} = \Omega^2\, \widetilde{\delta}_{\mu}^{\alpha}$ and the general relation,
\eqb
 g^{\mu \alpha} \frac{\delta}{\delta g^{\alpha \nu}} =
 \tg^{\mu \alpha} \frac{\delta}{\delta \tg^{\alpha \nu}} \, ,
\label{eq-app.four-dif.rel}
\eqe
are used to get the first term of the right-hand side of eq.~\eqref{eq-app.four-stress.2}. 
The trace $g^{\alpha \beta}\tmn{\alpha \beta}$ is usually called the conformal anomaly or the trace anomaly, and it is well known that the divergent part $W_{\rm div}$ gives the conformal anomaly as (see \S 6.3 in~\cite{ref-qft} for example)
\eqb
 g^{\alpha \beta}\tmn{\alpha \beta} =
 \frac{\Omega}{\sqrt{-g}}\, \frac{\delta W_{\rm div}}{\delta\Omega} \, .
\eqe
Hence substituting this expression of the conformal anomaly into eq.~\eqref{eq-app.four-stress.2} and using eq.~\eqref{eq-app.four-dif.rel} and eq.~\eqref{eq-app.four-action} with replacing $W_{\rm ren}$ by $W_{\rm div}$, we obtain
\eqb
 \tmn{\mu \nu}
 = \frac{1}{\Omega^2} \ttmn{\mu \nu}
 - \frac{2}{\sqrt{-g}}\, \frac{\delta W_{\rm div}}{\delta g^{\mu \nu}}
 + \frac{2\,\Omega^2}{\sqrt{-g}}\, \frac{\delta \tW_{\rm div}}{\delta \tg^{\mu \nu}} \, .
\eqe

The divergent part $W_{\rm div}$ can be evaluated from the Green function of the matter field $\phi$. We do not follow the details of the calculation of $W_{\rm div}$, but quote only the result for $\tmn{\mu \nu}$ for the conformally coupled massless scalar field $\phi$ on the spacetime of metric \eqref{eq-app.four-conf.metric} (see~\cite{ref-expect.stress.4D} or \S 6.2 and \S 6.3 in~\cite{ref-qft} for detail),
\eqb
 \tmn{\mu \nu}
 = \frac{1}{\Omega^2}\, \ttmn{\mu \nu}
 - \frac{1}{2880 \pi^2}\left( \frac{1}{6}\,X_{\mu \nu} - Y_{\mu \nu} \right)
 + 
   \frac{1}{2880 \pi^2\Omega^{2}}
   \left( \frac{1}{6}\, \widetilde{X}_{\mu \nu} - \widetilde{Y}_{\mu \nu} \right) \, ,
\label{eq-app.four-stress}
\eqe
where
\seqb
\label{eq-app.four-XY}
\eqab
 X_{\mu \nu} &\defeq&
    2\, \nabla_{\mu} \nabla_{\nu} {\cal R} - 2\,g_{\mu \nu}\, \Box {\cal R}
  + \frac{1}{2}\,{\cal R}^2\,g_{\mu \nu} - 2\,{\cal R}\, R_{\mu \nu} \, , \\
 Y_{\mu \nu} &\defeq&
  - R_{\mu}^{\alpha} R_{\alpha \nu} + \frac{2}{3}\, {\cal R}\, R_{\mu \nu}
  + \frac{1}{2}\, R_{\alpha \beta} R^{\alpha \beta}\, g_{\mu \nu}
  - \frac{1}{4} {\cal R}^2\,g_{\mu \nu} \, ,
\eqae
\seqe
where $R_{\mu \nu}$ and ${\cal R}$ are the Ricci tensor and scalar with respect to $g_{\mu \nu}$ respectively, and $\widetilde{X}_{\mu \nu}$ and $\widetilde{Y}_{\mu \nu}$ are defined similarly with respect to the metric $\tg_{\mu \nu}$. 
Equation \eqref{eq-app.four-stress} is the generalisation of eq.~(6.141) in~\cite{ref-qft} to the general conformal spacetimes of metric \eqref{eq-app.four-conf.metric}.



\end{document}